\documentclass{iopart}       
\usepackage{iopams}
\usepackage{setstack}
\usepackage[dvips]{graphics}

\def\ket#1{\vert#1\rangle}
\def\avec{{\bf a}}
\def\braket#1#2{\langle#1\vert#2\rangle}
\def\ketbra#1#2{\vert#1\rangle\langle#2\vert}
\def\bvec{{\bf b}}
\def\jvec{{\bf j}}
\def\Jvec{{\bf J}}
\def\nvec{{\bf n}}
\def\qvec{{\bf q}}
\def\zbar{{\bar z}}
\def\zhat{\hat{\bf z}}
\def\zerovec{{\bf 0}}

\def\Atilde{{\tilde A}}
\def\Btilde{{\tilde B}}
\def\Ttilde{{\tilde T}}
\def\Reals{{\mathbb R}}
\def\Realsdot{{\dot{\mathbb R}}}
\def\Complexes{{\mathbb C}}
\def\Complexesdot{{\dot{\mathbb C}}}
\def\Liealgebra#1{{\mathfrak #1}}
\def\TstarSU2dot{{\dot{{\displaystyle T^*}SU(2)}}}
\def\Sigmadot{{\dot\Sigma}}

\def\Shapespace{{\cal S}}
\def\Shapespacedot{{\dot\Shapespace}}

\def\Ldot{{\dot L}}
\def\Hspace{{\cal H}}
\def\Lspace{{\cal L}}
\def\Ghat{{\hat G}}

\def\Qdot{{\dot Q}}

\def\ket#1{\vert#1\rangle}

\def\braket#1#2{\langle#1\vert#2\rangle}
\def\ketbra#1#2{\vert#1\rangle\langle#2\vert}
\def\matrixelement#1#2#3{\langle#1\vert#2\vert#3\rangle}

\begin{document}
\title{Symplectic and Semiclassical Aspects of the Schl\"afli Identity}
\author{Austin Hedeman, Eugene Kur,
and Robert G. Littlejohn}
\address{Department of Physics, University of
California, Berkeley, California 94720 USA}
\author{Hal M. Haggard}
\address{Aix-Marseille Universit\'e and Universit\'e de Toulon, 
CPT-CNRS, Luminy, F-13288 Marseille and Physics program, Bard 
College, Annandale-on-Hudson, NY 12504, USA}

\ead{robert@wigner.berkeley.edu}

\begin{abstract}
  The Schl\"afli identity, which is important in Regge calculus and
  loop quantum gravity, is examined from a symplectic and
  semiclassical standpoint in the special case of flat, 3-dimensional
  space.  In this case a proof is given, based on symplectic geometry.
  A series of symplectic and Lagrangian manifolds related to the
  Schl\"afli identity, including several versions of a Lagrangian
  manifold of tetrahedra, are discussed.  Semiclassical
  interpretations of the various steps are provided.  Possible
  generalizations to 3-dimensional spaces of constant (nonzero)
  curvature, involving Poisson-Lie groups and $q$-deformed spin
  networks, are discussed.
\end{abstract}

\pacs{04.60.Nc, 04.60.Pp, 02.40.Yy, 03.65.Sq}


\section{Introduction}

The Schl\"afli identity, which is familiar in applications of Regge
calculus in general relativity (Regge 1961; Misner, Thorne and Wheeler
1973; Regge and Williams 2000), is a differential relation connecting
the volume of a polyhedron in an $n$-dimensional space of constant
curvature with the $(n-2)$-volumes and dihedral angles of its
$(n-2)$-dimensional faces.  In this article we deal with the special
case of a tetrahedron in Euclidean $\Reals^3$, for which the identity
itself is given by (\ref{Schlafli2}) below.  In this case we provide
an apparently new proof of the Schl\"afli identity, one based on
symplectic geometry.  We also discuss some geometrical constructions
related to the proof, including a pair of symplectic reductions that
take us from a 48-dimensional symplectic manifold in which the proof
is set to a 12-dimensional symplectic manifold in which the space of
tetrahedra is realized as a Lagrangian submanifold.  The inspiration
for our proof comes from a semiclassical or asymptotic analysis of the
Wigner $6j$-symbol (Wigner 1959, Edmonds 1960, Ponzano and Regge 1968,
Schulten and Gordon 1975ab, Biedenharn and Louck 1981, Roberts 1999,
Taylor and Woodward 2005, Aquilanti \etal 2012). As our analysis
proceeds we provide semiclassical interpretations of many of the
steps.

The Schl\"afli identity is useful for obtaining formulas for the
volume of polyhedra in spaces of constant curvature.  Schl\"afli's
(1858) original derivation concerned spaces of positive curvature (we
note that several of the citations to this article in the recent
literature are incorrect).  The result was generalized to spaces of
negative curvature by Sforza (1907) and new proofs given by Kneser
(1936).  More recent treatments include Milnor (1994),
Alekseevskij, Vinberg and Solodovnikov (1993) and Yakut, Savas and
Kader (2009).

Another approach to the Schl¨afli identity has been explored recently
(Rivin and Schlenker 2000, Souam 2004). It begins with a formula which
is valid for an arbitrary hypersurface embedded in an Einstein
manifold (a manifold of constant curvature whose metric satisfies the
Einstein equations with a cosmological constant). The formula relates
the variation of the volume enclosed by the hypersurface to variations
of the extrinsic curvature and induced metric on the hypersurface. We
have shown that the same formula can be obtained by varying the
Einstein-Hilbert action (the integral of the curvature scalar)
evaluated on the enclosed region and relating that variation to a
surface integral over the boundary using manipulations similar to
those employed in the ADM formalism in general relativity (Misner,
Thorne and Wheeler 1973, Thiemann 2007).  For a polyhedron, the
extrinsic curvature is concentrated in the manner of a delta function
on the codimension two faces of the polyhedron, where its integral is
related to the dihedral angles. When applied to the polyhedron, the
formula thus reduces to the Schl\"afli identity.

Although our (symplectic) proof of the Schl\"afli identity stands on
its own, the applications that we have in mind are related to Regge
calculus, an approach to discretizing general relativity, and to
loop quantum gravity, which also involves a discretization of the
degrees of freedom of the gravitational field.  See, for example,
Barrett and Steele (2003), Livine and Oriti (2003), Dittrich, Freidel
and Speziale (2007), and Bahr and Dittrich (2009, 2010), where the
Schl\"afli identity plays a crucial role.  See also Carfora and
Marzuoli (2012) for a modern, comprehensive review of simplicial
methods in quantum gravity and other fields, including the role of
state sums and their regularizations.  We suspect that our
symplectic approach to the Schl\"afli identity may be especially
relevant in loop quantum gravity, where spin networks such as the
$6j$-symbol play an important role and where the classical phase space
(Freidel and Speziale 2010, Livine and Tambornino 2011) has been
identified with the same symplectic manifolds that appear in our
analysis.

We begin in Sec.~\ref{LMoftetrahedra} by explaining the shape space of
tetrahedra and by presenting an {\it ad hoc} construction of a
symplectic manifold and a submanifold thereof that can be identified
with the space of tetrahedra.  This submanifold is Lagrangian by
virtue of the Schl\"afli identity, with a generating function
(\ref{Sdef}) that we call the ``Ponzano-Regge phase.''  In
Sec.~\ref{PRphase} we present an integral representation of the
Ponzano-Regge phase, essentially a derivation of the phase of the
asymptotic expression for the Wigner $6j$-symbol, following the method
of Roberts (1999).  In Sec.~\ref{proof} we present a proof of the
Schl\"afli identity using this integral representation.  The method of
proof involves integrating the symplectic form over a certain surface
in phase space and then using Stokes' theorem, in a manner common to
several basic proofs in classical mechanics (Arnold 1989).  Then in
Secs.~\ref{firstreduction} and \ref{secondreduction} we carry out a
sequence of two symplectic reductions on the symplectic manifold in
which the proof of the Schl\"afli identity is set, recovering at the
end the symplectic manifold and Lagrangian submanifold of tetrahedra
that we started with in Sec.~\ref{LMoftetrahedra}.  In
Sec.~\ref{conclusions} we present various remarks and conclusions,
including a outline of some of the features of the generalization of
this work to the $q$-deformed $6j$-symbol and the Schl\"afli identity
in spaces of constant, nonzero curvature.

\section{The Shape Space and Lagrangian Manifold of Tetrahedra}
\label{LMoftetrahedra}

\subsection{Tetrahedra and Their Shapes}
\label{shapes}

A tetrahedron may be defined as a subset of Euclidean $\Reals^3$ in
terms of an ordered set of four points, which are the vertices.  The
six edges and their lengths are defined in terms of the vertices.  As
special cases we allow the tetrahedron to be flat (the four vertices
may lie in a plane) or of lower dimensionality (the vertices may lie
in a line or they may all coincide).  Any or all of the vertices are
allowed to coincide, so that some or all of the edge lengths may be
zero.

By this definition the space of tetrahedra is $(\Reals^3)^4$.  If we
consider two tetrahedra equivalent that are related by translations,
then the space reduces to $(\Reals^3)^3$, in which the three vectors
can be taken as the edge vectors emanating from a given vertex.  If we
expand the equivalence classes to include proper rotations, then the
space of tetrahedra reduces to what we will call the ``shape space of
tetrahedra,''
	\begin{equation}
	\Shapespace = \frac{(\Reals^3)^3}{SO(3)} \cong \Reals^6,
	\label{shapespace}
	\end{equation}
where the action of $SO(3)$ is the diagonal action on all three copies
of $\Reals^3$ (that is, it is a rigid, proper rotation of the
tetrahedron) and where $\cong$ means ``is diffeomorphic to.''  This is
shown by Narasimhan and Ramadas (1979) and discussed further by
Littlejohn and Reinsch (1995).  In the following two tetrahedra will
be considered to have the same shape if they are related by a translation
and a proper rotation.

Then it turns out that the space of flat tetrahedra (those whose
vertices lie in a plane in $\Reals^3$) constitute a subspace
$\Reals^5\subset\Shapespace\cong\Reals^6$, which we will call the ``shape
space of flat tetrahedra.''  If we define the volume $V$ of the
tetrahedron as $1/6$ the triple product of three vectors emanating
from a given vertex, then the subspace $\Reals^5$ of flat tetrahedra
divides shape space $\Shapespace$ into three subsets, those on which
$V>0$, $V<0$ and $V=0$ (the last being the subspace $\Reals^5$ of
flat tetrahedra itself).  Furthermore, performing a spatial inversion
of a tetrahedron in $\Reals^3$ causes the point of $\Shapespace$ to be
reflected in the hyperplane $\Reals^5$ of flat tetrahedra.  In
particular, the shape of a tetrahedron is invariant under spatial
inversion if and only if it is flat.

It is also shown in the references cited that the six edge lengths of
the tetrahedron form a coordinate system on the regions $V\ge0$ and
$V\le0$ of $\Shapespace$, that is, there is a one-to-one map from these
regions of $\Shapespace$ to a region of the six dimensional space with
coordinates $J_r$, $r=1,\ldots,6$, where $J_r$ is the length of edge
$r$.  This map is not onto, however, since there are values of the
$J_r$ that do not correspond to any tetrahedron.  In the first place
these lengths obviously must satisfy $J_r\ge0$ and the four triangle
inequalities for the four faces of the tetrahedron; in addition, there
is a further requirement that the faces can be assembled into a
tetrahedron.  All these conditions can be expressed in terms of the
minors of the Cayley-Menger determinant (Ponzano and Regge 1968) or of
an associated Gram matrix (Littlejohn and Yu 2009).  Spatial inversion
maps a tetrahedron with volume $V$ into one with volume $-V$, without
changing the edge lengths $J_r$; therefore, given edge lengths such
that a tetrahedron exists, the shape of the tetrahedron is determined
to within a spatial inversion (hence uniquely, for a flat
tetrahedron).

We now define the dihedral angle $\psi_r$ associated with edge $r$ for
a tetrahedron of a given shape, first for the case $V\ge0$.  A given
edge is the intersection of two adjacent faces; the dihedral angle is
not defined unless the areas of the two faces are nonzero, so we
assume this.  (If a face has zero area, then we will refer to it as
``degenerate.'')  Then we define the dihedral angle $\psi_r$ as the
angle between the outward pointing normals to the two faces.  This
gives $0\le\psi_r\le\pi$.  Finally, if $V<0$, we define $\psi_r$ as
the negative of the angle $\psi_r$ for the spatially inverted shape
(which has $V>0$).  With these conventions, the dihedral angles lie in
the range $-\pi<\psi_r\le\pi$, and all dihedral angles change sign
under spatial inversion, modulo $2\pi$.

The subset of shape space on which one or more dihedral angles are not
defined consists of tetrahedra with one or more degenerate faces.  All
such tetrahedra have zero volume, so they form a subset of the shape
space $\Reals^5$ of flat tetrahedra.  This subset has codimension 1
inside $\Reals^5$ (more precisely, it is the union of smooth
manifolds, of which the maximum dimensionality is 4).  If the dihedral
angles $\psi_r$ are defined for a flat tetrahedron, then they are
either 0 or $\pi$, and are constant inside connected regions of the
space $\Reals^5$ of flat tetrahedra; these regions are separated by
the codimension 1 subset upon which some face is degenerate.  We will
denote the subset of shape space $\Shapespace$ upon which the dihedral
angles are defined by $\Shapespacedot$; it is $\Shapespace$ minus the
tetrahedra with one or more degenerate faces.

Our definition of the dihedral angles differs from the usual one used
in discussions of the Schl\"afli identity, which is the absolute value
of the definition given here. Our definition has the advantage that
the dihedral angles are smooth functions along a smooth curve crossing
the subspace $\Reals^5$ of flat tetrahedra, modulo $2\pi$, if we avoid
shapes with degenerate faces.  More precisely, all dihedral angles are
either 0 or $\pi$ on the subspace $\Reals^5$ of flat tetrahedra, as
long as we avoid degenerate faces; as we pass from the region $V>0$ to
the region $V<0$ through this subspace, an angle which is 0 on the
flat tetrahedron passes smoothly from positive to negative values,
while an angle that is $\pi$ on the flat tetrahedron jumps
discontinuously from $+\pi$ to $-\pi$.  In both cases, the
differential $d\psi_r$ is smooth.  As an example of such a motion we
may rotate two adjacent faces relative to one another about their
common edge, so that one face passes through the plane of the other.
Along this motion the lengths of all the edges are constant except for
the one opposite the edge common to the two faces.

We remark that the inclusion of negative angles in the definition of
the dihedral angles is merely a convenience in the case of the
asymptotics of the $6j$-symbol, but such an extension to a full range
of $2\pi$ in the dihedral angles is necessary for more complex spin
networks, such as the $9j$-symbol.  The definition of dihedral angles
given here is equivalent to the general definition given by Haggard
and Littlejohn (2010).  In addition, negative dihedral angles emerge
naturally at the end of the sequence of symplectic reductions carried
out in this paper (see Sec.~\ref{secondreduction}).

If a tetrahedron has one or more edge lengths that are zero (that is,
if two or more vertices coincide), then two or more faces are
degenerate; these tetrahedra form a set of shapes that is a subset of
the tetrahedra with degenerate faces.  It has codimension 2 inside
the space $\Reals^5$ of flat tetrahedra.

\subsection{Lagrangian Interpretation of the Schl\"afli Identity}
\label{lagrangianinterp}

For the time being we concentrate on the region $V>0$, where the
dihedral angles $\psi_r$ are functions of the edge lengths $J_r$.
Since the $\psi_r$ do not change if the edge lengths are scaled by
some positive factor, Euler's theorem on homogeneous functions implies
    \begin{equation}
      \sum_r J_r \frac{\partial \psi_s}{\partial J_r}=0.
      \label{Euler}
    \end{equation}
The Schl\"afli identity has a similar appearance; it is
    \begin{equation}
      \sum_r J_r \frac{\partial \psi_r}{\partial J_s}=0,
      \label{Schlafli1}
      \end{equation}
which, after multiplying by $dJ_s$ and summing over $s$, can be written
    \begin{equation}
      \sum_r J_r \,d\psi_r = 0.
      \label{Schlafli2}
      \end{equation}
See Luo (2008) for other identities involving the matrix $\partial
\psi_r/\partial J_s$, including the case of tetrahedra in spaces of
constant (nonzero) curvature.

If we differentiate (\ref{Schlafli1}) with respect to $J_k$ and
antisymmetrize in $k$ and $s$, we obtain
    \begin{equation}
	\frac{\partial \psi_k}{\partial J_s} =
	\frac{\partial \psi_s}{\partial J_k},
	\label{symmetricJac}
	\end{equation}
that is, the matrix $\partial \psi_k/\partial J_s$ is symmetric.  Thus
the Schl\"afli identity (\ref{Schlafli1}) implies the Euler identity
(\ref{Euler}).  It also implies	
	\begin{equation}
	\psi_r = \frac{\partial S}{\partial J_r},
	\label{genfunrel}
	\end{equation}
where $S$ is given by
	\begin{equation}
	S = \sum_r J_r\,\psi_r.
	\label{Sdef}
	\end{equation}
We refer to $S$ as the ``Ponzano-Regge phase'' since
it appears in the factor $\cos(S+\pi/4)$ in the asymptotic formula for
the Wigner $6j$-symbol due to Ponzano and Regge (1968).  

\begin{figure}[htb]
\begin{center}
\scalebox{0.43}{\includegraphics{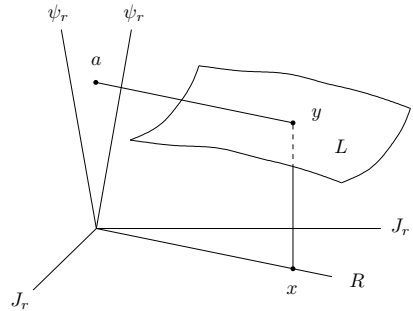}}
\end{center}
\caption[rps]{\label{schlafliL} The tetrahedra form a Lagrangian
manifold in the 12-dimensional space with coordinates $J_r$ and
$\psi_r$, $r=1,\ldots,6$.}
\end{figure}

These relations imply that the space of tetrahedra is a Lagrangian
submanifold of a 12-dimensional space with coordinates
$(J_1,\ldots,J_6,\psi_1,\ldots,\psi_6)$, of which a schematic
illustration is given in Fig.~\ref{schlafliL}.  The
6-dimensional manifold $L$ is the graph of the functions
(\ref{genfunrel}); because it is the graph of a gradient, $L$ is
Lagrangian with respect to the symplectic form,
	\begin{equation}
	\sum_r dJ_r \wedge d\psi_r = d\sum_r J_r \, d\psi_r.
	\label{schlafliomega}
	\end{equation}
The figure shows a point $x$ of the 6-dimensional space of edge
lengths (call it ``$J$-space'') and the vertical manifold of constant
edge lengths above it which intersects $L$ in the point $y$, which in
turn projects onto the space of angles (call it ``$\psi$-space'') at
point $a$.  Due to the homogeneity of the functions
$\psi_r(J_1,\ldots,J_6)$, the point $a$ does not move if $x$ is moved
along the radial line $R$ in $J$-space, that is, if the edge lengths
are scaled by a common factor.  This means that the line $a$--$y$ is
everywhere tangent to $L$, and that $L$ is vertical in one dimension
over $\psi$-space.  Manifold $L$, which is 6-dimensional and which has
a nonsingular projection onto $J$-space, has only a 5-dimensional
projection onto $\psi$-space.  That is, points of $\psi$-space
that can be realized as dihedral angles of a tetrahedron are first
order caustic points of the projection from $L$.

The construction given has applied in the region $V>0$.  The same
construction works in the region $V<0$, providing a second branch to
the manifold illustrated in Fig.~\ref{schlafliL}, obtained from the
first by $\psi_r \mapsto -\psi_r$.  On the subset $V=0$, excluding
shapes with degenerate faces, the main problem is the angles $\psi_r$
that jump discontinuously from $\pi$ to $-\pi$.  We can avoid this by
speaking of a trivial $T^6$-bundle over $\Shapespacedot$. Then what we
are calling the functions $\psi_r(J_1,\ldots,J_6)$ becomes a smooth
section of this bundle, that is, the surface $L$, extended to flat
shapes with nondegenerate faces. Then the differential forms $d\psi_r$
and the combination seen in (\ref{Schlafli2}) are well defined
everywhere in the bundle, and the Schl\"afli identity in the form
(\ref{Schlafli2}) holds everywhere on $L$.  Also, the Ponzano-Regge
phase $S$, which involves the angles $\psi_r$ themselves (not their
differentials), is a function on $L$; but it is discontinuous at the
flat shapes.

In our next step we exploit a certain integral representation for the
Ponzano-Regge phase $S$.  It turns out that $S$ can be expressed as a
line integral of a symplectic 1-form in a certain 48-dimensional
symplectic manifold along a path running along one 24-dimensional
Lagrangian submanifold and then back along another, in the type of
geometry that arises in the semiclassical analysis of scalar products
$\braket{A}{B}$ in quantum mechanics.  This is explained in
Sec.~\ref{PRphase}.  This representation is then used in
Sec.~\ref{proof} to prove the Schl\"afli identity.  Then in subsequent
sections it is shown that the 12-dimensional symplectic manifold
illustrated in Fig.~\ref{schlafliL} (``$J$--$\psi$-space'' or the
$T^6$ bundle just described) can be obtained from the 48-dimensional
one in which the integral representation is expressed by means of two
symplectic reductions.  This process reveals information about the
symplectic manifold in Fig.~\ref{schlafliL} that is not apparent from
the presentation so far.

\section{An Integral Representation of the Ponzano-Regge Phase}
\label{PRphase}

In this section we present a derivation of the Ponzano-Regge phase as
the principal contribution to the phase of the asymptotic expression
for the Wigner $6j$-symbol.  The derivation is a reformulation of that
given by Roberts (1999), which as far as we know is the most
symmetrical and elegant available.  We express it, however, in a
somewhat different geometrical language than Roberts, using real
Lagrangian manifolds that are parameterized as level sets of various
momentum maps and an arbitrary real representation of the wave
functions, rather than a complex or coherent state representation. We
invoke the semiclassical treatment of the $6j$-symbol merely to
motivate the geometry behind the representation of the Ponzano-Regge
phase as a certain line integral along certain Lagrangian manifolds;
this is what we will need for our proof of the Schl\"afli identity in
Sec.~\ref{proof}.  For the purposes of this paper the manifolds need
not be quantized.

Our brief discussion of the semiclassical mechanics of the $6j$-symbol
involves what we call the ``$12j$-model'' of the $6j$-symbol, which is
due to Roberts (1999) and which is explained in terms of spin networks
in Sec.~3 of Aquilanti \etal (2012).  In this model, the $6j$-symbol is
expressed as the scalar product $\braket{A}{B}$ of two vectors
$\ket{A}$ and $\ket{B}$ in a certain Hilbert space, where the labels
$A$ and $B$ stand for two complete sets of commuting observables of
which the vectors are eigenvectors, with certain normalization and
phase conventions. The two complete sets involve angular momentum
operators whose components do not commute except on the subspace where
the angular momenta vanish, which is precisely the subspace in
terms of which the $6j$-symbol is defined.  These two sets of
commuting operators correspond classically to two sets of Poisson
commuting functions on a certain phase space.  Again, the sets involve
various (now classical) angular momenta whose components do not
Poisson commute except on the level set upon which the angular momenta
vanish, which are precisely the Lagrangian manifolds relevant to the
semiclassical evaluation of the scalar product $\braket{A}{B}$.

\subsection{Phase spaces}
\label{phasespaces}

We now explain the phase space in which these manifolds live.  The
notation is the same as in Aquilanti \etal (2007, 2012), with minor
modifications which are noted.  We begin with the phase space or
symplectic manifold $\Phi$ of a 2-dimensional, isotropic harmonic
oscillator with unit mass and frequency.  This is a classical version
of the harmonic oscillator system used by Schwinger (1952) and
Bargmann (1962) in their treatment of the representation theory of
$SU(2)$, which anticipated many of the features of the modern theory
of geometric quantization (Kirillov 1976, Simms and Woodhouse 1977,
Bates and Weinstein 1997, Echeverr\'\i a-Enr\'\i quez \etal 1999).
Coordinates on $\Phi$ are $(x_1,x_2,p_1,p_2)$, so that as a symplectic
manifold $\Phi=(\Reals^4,dp\wedge dx)$, where $dp\wedge dx$ means
$\sum_\mu dp_\mu \wedge dx_\mu$.  We use indices $\mu$, $\nu$, etc to
run over 1,2, indexing the two harmonic oscillators; we often suppress
these indices with an implied summation.  We introduce complex
coordinates $z_\mu = (x_\mu+ip_\mu)/\sqrt{2}$, $\zbar_\mu
=(x_\mu-ip_\mu) /\sqrt{2}$ on $\Phi$, so that $\Phi$ can be seen as
$(\Complexes^2,idz^\dagger \wedge dz)$, where $z$ (without the $\mu$
subscript) is seen as a column vector or spinor $(z_1,z_2)^T$, where
$z^\dagger$ is seen as a row spinor $(\zbar_1,\zbar_2)$ and where
again a summation over $\mu$ is implied.  We denote the symplectic
2-form on $\Phi$ by $\omega = d\theta$, where $\theta$ can be taken to
be $iz^\dagger \, dz$.  Although $\theta$ is complex, we will be only
interested in integrals of it along closed loops, for which the
integrals are real.

The phase space $\Phi$ has several functions defined on it, including
         \begin{equation}
           I = \frac{1}{2} z^\dagger z, \qquad
           J_i = \frac{1}{2} z^\dagger \sigma_i z,
           \label{IJdef}
           \end{equation}
where $\sigma_i$, $i=1,2,3$ are the Pauli matrices and where spinor
contractions are implied.  We note that $I=H/2$, where $H$ is the
harmonic oscillator $(1/2)\sum_\mu (p_\mu^2 + x_\mu^2)$.  These
functions satisfy $\Jvec^2 = I^2$, where we use bold face $\Jvec$ for
the 3-vector with components $J_i$.  There are also the Poisson
bracket relations $\{I,J_i\}=0$, $\{J_i,J_j\} = \epsilon_{ijk}\,J_k$.
The definition of $J_i$ in (\ref{IJdef}) implies a map
$\pi_H:\Complexes^2 \to
\Reals^3$, which we call the Hopf projection because when restricted
to the 3-sphere $I={\rm const} > 0$ it is the projection map of the
Hopf fibration of $S^3$ over $S^2$ (Cushman and Bates 1997, Holm
2011).  Here $\Reals^3$ is the space with coordinates $\Jvec$, which
we call ``angular momentum space'' (it is the dual of the Lie algebra
of $SU(2)$).

We construct larger symplectic manifolds by taking products of
$\Phi$.  First we construct the symplectic manifold
      \begin{equation}
        \Phi_{2j} = (\Complexes^2 \times \Complexes^2, 
        idz^\dagger\wedge dz + i
        dz^{\prime\dagger} \wedge dz'),
        \label{Phi2jdef}
      \end{equation}
where spinors $z$ and $z'$ are coordinates in the first and second
copies of $\Complexes^2$.  For brevity we will write this as
$\Phi_{2j} = \Phi \times \Phi$ or $\Complexes^2 \times
\Complexes^2$, with  the symplectic form in (\ref{Phi2jdef})
understood.  We introduce primed versions of (\ref{IJdef}) to define
$I'$ and $\Jvec'$, so that $\Phi_{2j}$ has functions $I$, $\Jvec$,
$I'$ and $\Jvec'$ defined on it.

On $\Phi_{2j}$ $I$ and $I'$ generate the $U(1)$ actions, $z\mapsto
e^{-i\alpha/2}z$ and $z'\mapsto e^{-i\alpha/2}z'$, respectively, where
$\alpha$ is the angle conjugate to the Hamiltonian function $I$ or
$I'$.  Similarly, $\Jvec$ and $\Jvec'$ generate $SU(2)$ actions, that
is, Hamiltonian function $H=\nvec\cdot\Jvec$, where $\nvec$ is a unit
vector, generates the action,
	\begin{equation}
	z \mapsto u(\nvec,\alpha)z, \qquad z'\mapsto z',
	\label{Jaction}
	\end{equation}
where $\alpha$ is the angle conjugate to $H$ and where
        \begin{equation}
          u(\nvec,\alpha)=e^{-i(\alpha/2)(\nvec\cdot\bsigma)}
          \label{uaxisangle}
        \end{equation}
represents an element of $SU(2)$ in axis-angle form.  Similarly,
$H=\nvec\cdot\Jvec'$ generates the action
	\begin{equation}
	z\mapsto z, \qquad z' \mapsto u(\nvec,\alpha)z'.
	\label{Jprimeaction}
	\end{equation}

Finally we define the symplectic manifold
       \begin{equation}
         \Phi_{12j} = (\Phi_{2j})^6,
         \label{Phi12jdef}
       \end{equation}
with coordinates $z_r$ and $z'_r$, $r=1,\ldots,6$, where $r$ labels
the factors in (\ref{Phi12jdef}).  The symplectic 1-form and 2-form
on $\Phi_{12j}$ are
       \begin{equation}
         \theta=\sum_r iz^\dagger_r\, dz_r + iz^{\prime\dagger}_r \,
         dz'_r,
         \label{theta12j}
       \end{equation}
and
       \begin{equation}
         \omega=d\theta=\sum_r idz^\dagger_r \wedge dz_r
         + idz^{\prime\dagger}_r \wedge dz'_r.
         \label{omega12j}
       \end{equation}
By adding an $r$-subscript to the functions defined in (\ref{IJdef}),
with or without a prime, we obtain functions $I_r$, $I'_r$, $\Jvec_r$,
$\Jvec'_r$, $r=1,\ldots,6$ on $\Phi_{12j}$.  By extending $\pi_H$ to
all twelve copies of $\Complexes^2$ we obtain a projection
$\pi_H:\Phi_{12j} \to (\Reals^3)^{12}$, where the latter space is the
angular momentum space for all twelve angular momenta, $\Jvec_r$,
$\Jvec'_r$, $r=1,\ldots,6$.

\subsection{The $A$- and $B$-manifolds}
\label{ABmanifolds} 

Now we introduce two Lagrangian manifolds in $\Phi_{12j}$, specified
as the level sets of collections of functions that Poisson commute on
the manifolds.  We call these the $A$- and $B$-manifolds.  The
functions defining the $A$-manifold and their contour values are given
by
       \begin{equation}
         \fl\hbox{\rm $A$-set} =
         \left(\begin{array}{cccccccccc}
             I_1 & \cdots & I_6 & I'_1 & \cdots & I'_6 &
             \Jvec_{123} & \Jvec_{1'5'6} & \Jvec_{2'6'4} &
             \Jvec_{3'4'5} \\
             J_1 & \cdots & J_6 & J_1 & \cdots & J_6 &
             \zerovec & \zerovec & \zerovec & \zerovec
             \end{array}\right),
           \label{Aset}
         \end{equation}
where the first row contains the functions and the second row contains
the contour values, and where $\Jvec_{123} = \Jvec_1+\Jvec_2+\Jvec_3$,
$\Jvec_{1'5'6} = \Jvec'_1 + \Jvec'_5 + \Jvec_6$, $\Jvec_{2'6'4} =
\Jvec'_2 + \Jvec'_6 + \Jvec_4$ and $\Jvec_{3'4'5}=\Jvec'_3 + \Jvec'_4
+ \Jvec_5$.  Notice that the contour values of the primed $I'_r$ are
the same as those of the unprimed $I_r$, that is, $J_r$.  It is
assumed in the following that the $J_r$ are given
numbers such that a tetrahedron exists with edge lengths $J_r$, and
that the faces are nondegenerate so that the dihedral angles $\psi_r$
are defined.  In particular, this means that $J_r>0$, $r=1,\ldots,6$.
There are 24 functions in the $A$-list (six $I$'s, six
$I'$'s and $4\times3=12$ components of angular momenta).  For the
given values of $J_r$ these functions are independent and define a
24-dimensional submanifold of $\Phi_{12j}$ (the $A$-manifold).  Also,
since the functions Poisson commute with each other on the
$A$-manifold, it is Lagrangian (see the discussion in Sec.~4.3 of
Aquilanti \etal (2012)).

The vanishing of the four sums of three angular momenta indicated by
(\ref{Aset}) represents four triangle conditions among the twelve
angular momenta and defines four triangles in a single copy of
$\Reals^3$, if all twelve vectors are plotted in this space.  The
choice of the particular sets of three angular momenta to form
triangles is governed by the operators of which the vector $\ket{A}$
is an eigenvector, as explained in Sec.~3 of Aquilanti \etal (2007).
It is also a consequence of the rules explained in Haggard and
Littlejohn (2010) for translating a spin network into a stationary
phase condition involving collections of angular momenta.

The functions defining the $B$-manifold and their contour values are
given by
       \begin{equation}
         \fl\hbox{\rm $B$-set} =
         \left(\begin{array}{ccccccccc}
              I_1 & \cdots & I_6 & I'_1 & \cdots & I'_6 &
              \Jvec_{11'} & \ldots & \Jvec_{66'} \\
              J_1 & \cdots & J_6 & J_1 & \cdots & J_6 &
              \zerovec & \cdots & \zerovec
            \end{array}\right)
          \label{Bset}
        \end{equation}
where $\Jvec_{11'} = \Jvec_1+\Jvec'_1$ etc.  Notice that the
contour values of the functions $I_r$ and $I'_r$ are the same as on
the $A$-manifold (\ref{Aset}).  Now there are twelve conditions on
$I_r$ and $I'_r$ and $3\times6=18$ components of angular momentum
vectors, or 30 conditions altogether; but these are not 
independent, since $\Jvec_r = -\Jvec'_r$ implies $I_r = I'_r$.  Thus
the conditions $I'_r=J_r$ are superfluous and can be dropped from the
list, giving 24 independent conditions.  Therefore the $B$-manifold is a
24-dimensional submanifold of $\Phi_{12j}$.  It is also Lagrangian, for the
same reason as the $A$-manifold.  We call the conditions $\Jvec_r +
\Jvec'_r=0$ the ``diangle conditions''; when they are satisfied,
$\Jvec_r$ and $\Jvec'_r$ are equal and opposite.

\begin{figure}[htb]
\begin{center}
\scalebox{0.43}{\includegraphics{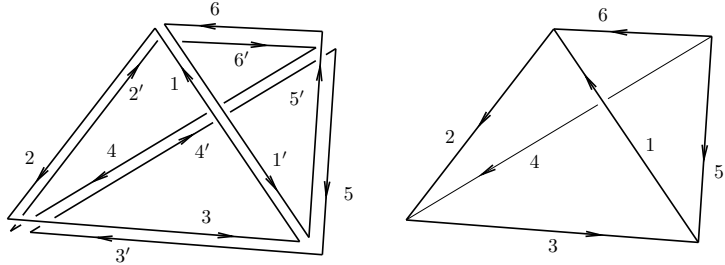}}
\end{center}
\caption[rps]{\label{tetrahedron} The four triangle conditions and the
six diangle conditions specify a tetrahedron, in which each edge is
represented by oppositely pointing vectors $\Jvec_r$ and $\Jvec'_r$ of
the same length.  This tetrahedron is illustrated on the left, with
the faces pulled away from one another slightly to make the opposite
vectors $\Jvec_r$, $\Jvec'_r$ more clear.  The symbols 1, 2, etc
represent vectors $\Jvec_1$, $\Jvec_2$, while $1'$, $2'$ etc represent
the vectors $\Jvec'_1$, $\Jvec'_2$, etc.  On the right the tetrahedron
is reassembled, with only the unprimed vectors shown.}
\end{figure}

\subsection{Intersections and Tetrahedra}

If the $A$- and $B$-manifolds intersect, then both the diangle and
triangle conditions on the twelve angular momenta hold, and the
angular momenta define a tetrahedron, as illustrated in
Fig.~\ref{tetrahedron}.  Each edge of the tetrahedron is represented
by a pair of oppositely pointing vectors $\Jvec_r$ and $\Jvec'_r$.
Conversely, if a tetrahedron exists with the given edge lengths $J_r$,
$r=1,\ldots,6$, then the $A$- and $B$-manifolds intersect.  Since we
are assuming that the values of the $J_r$ are such that a tetrahedron
does exist, we are assured that the $A$ and $B$ manifolds do
intersect.  

Generically two 24-dimensional manifolds in a 48-dimensional space
intersect in a discrete set of points (a 0-dimensional set), but the
intersections of the $A$ and $B$ manifolds (for the assumed values of
the $J_r$) are 15-dimensional.  The intersection has a nongeneric
dimensionality because of a ``common symmetry group'' between the $A$-
and $B$-manifolds, as explained in somewhat general terms in Sec.~6.2
of Aquilanti \etal (2012).  The basic idea is the following.  The
$A$-list of functions (\ref{Aset}) is the momentum map (Abraham and
Marsden 1978, Marsden and Ratiu 1999) of a group $G_a$ whose action on
$\Phi_{12j}$ is generated by the $A$-list of functions.  This group is
$G_a=U(1)^{12} \times SU(2)^4$, where the $U(1)$ factors are generated
by the $I_r$ and $I'_r$, and the $SU(2)$ factors are generated by the
four partial sums of angular momenta, $\Jvec_{123}$, $\Jvec_{1'5'6}$,
$\Jvec_{2'6'4}$ and $\Jvec_{3'4'5}$.  The action generated by one of
the $I_r$ or $I'_r$ is the changing of the overall phase of one of the
spinors $z_r$ or $z'_r$; for example, the Hamiltonian flow generated
by $I_r$ is
	\begin{equation}
	z_r \mapsto e^{-i\alpha/2}z_r,
	\label{Irflow}
	\end{equation}
where $\alpha$ is the parameter of the flow (it is the angle conjugate
to $I_r$).  Under this flow all unprimed spinors $z_s$ for $s\ne r$
and all primed spinors are not affected.  The action generated by
$I'_r$ is similar (it only affects $z'_r$).  The flow (\ref{Irflow})
is a motion along the fiber (the ``Hopf circle'') of the Hopf
fibration for the particular copy of $\Complexes^2$ on which $z_r$ is
a coordinate; the period of the circle is $\alpha=4\pi$.  All the
functions $I_r$, $I'_r$, $\Jvec_r$ and $\Jvec'_r$ are invariant along
the flow (\ref{Irflow}).  The action generated by one of the triangle
sums of angular momenta, $\Jvec_{123}$ for example, is the
multiplication of the selected spinors ($z_1$, $z_2$ and $z_3$ in this
case) by an element $u\in SU(2)$, which causes the corresponding
$\Jvec$ vectors ($\Jvec_1$, $\Jvec_2$ and $\Jvec_3$ in this case) to
be multiplied by $R(u) \in SO(3)$, where
	\begin{equation}
	R_{ij}(u) = \frac{1}{2} \tr(u^\dagger \sigma_i u \sigma_j).
	\label{Rijdef}
	\end{equation}
This is the standard projection from $SU(2)$ to $SO(3)$.  Under this
action, the other $\Jvec$ vectors are not affected, nor are
any of the $I_r$ or $I'_r$.  Similarly the $B$-list of functions is
the momentum map for a group $G_b=U(1)^{12} \times SU(2)^6$, where the
$U(1)$ factors are the same as in $G_a$, while each of the six copies
of $SU(2)$ is generated by $\Jvec_r+\Jvec'_r$ and affects a pair
$(z_r,z'_r)$ for a given $r$ (and thus the vectors $\Jvec_r$,
$\Jvec'_r$ are rotated by the same element $R(u)\in SO(3)$).  Because
the $A$- and $B$-manifolds are Lagrangian, these manifolds are not
only level sets of the corresponding momentum maps, but also the
orbits of the corresponding groups.

The group $G_c$ common to $G_a$ and $G_b$ has an action on
$\Phi_{12j}$ that consists of symplectic transformations common to the
$A$- and $B$-actions.  It is generated by $I_r$, $I'_r$ and by
	\begin{equation}
	\Jvec_{\rm tot} = \sum_r \Jvec_r + \Jvec'_r,
	\label{Jtotdef}
	\end{equation}
and thus it is the 15-dimensional group $G_c=U(1)^{12} \times SU(2)$,
where the single factor of $SU(2)$ multiplies all spinors $z_r$,
$z'_r$ by a common element of $SU(2)$.  The 15-dimensional orbits of
$G_c$ are common to both the $A$- and $B$-manifolds, and thus lie in
their intersection.  In fact the intersection is precisely
15-dimensional (for the given values of $J_r$).

\begin{figure}[htb]
\begin{center}
\scalebox{0.43}{\includegraphics{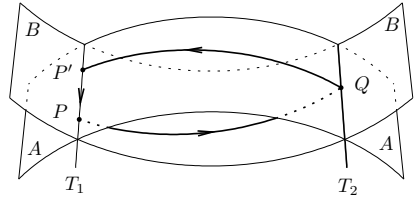}}
\end{center}
\caption[rps]{\label{ABintsect} The 24-dimensional $A$- and
$B$-manifolds intersect in two 15-dimensional submanifolds $T_1$ and
$T_2$, consisting of tetrahedra of a given shape and its image under
spatial inversion.  Also illustrated is the path $P\to Q\to P'\to P$
for the integral representation of the Ponzano-Regge phase.}
\end{figure}

If the volume of the tetrahedron is nonzero, however, then the
intersection of the $A$- and $B$-manifolds is not connected.  This is
because both the tetrahedron and its image under spatial inversion
simultaneously satisfy the $A$- and $B$-conditions, but the group
$G_c$ contains only proper rotations, and cannot map a tetrahedron of
nonzero volume into its image under inversion.  These facts lie behind
the schematic illustration of the $A$- and $B$-manifolds and their
intersections shown in Fig.~\ref{ABintsect}, where the two connected
pieces of the intersection are labeled $T_1$ and $T_2$ (for
``tetrahedra'', since these intersections are the places in
$\Phi_{12j}$ where the vectors $\Jvec$ and $\Jvec'$ form a
tetrahedron).  That is, Fig.~\ref{ABintsect} is drawn for the case
$V\ne0$, for which the manifolds $T_1$ and $T_2$ correspond to two
points of shape space $\Shapespace$, related by spatial inversion.  As
we move along $T_1$ or $T_2$, the only things that change are the
overall phases of the 12 spinors and the overall orientation of the
tetrahedron, neither of which changes the shape.

\subsection{A Path for the Ponzano-Regge Phase}
\label{pathPRphase}

The branches of the semiclassical approximation of a scalar product
such as $\braket{A}{B}$ correspond to intersections of the Lagrangian
manifolds corresponding to vectors $\ket{A}$ and $\ket{B}$, and the
relative phase between two branches is the integral of the symplectic
1-form along a path starting on one intersection, running along one of
the Lagrangian manifolds to the other intersection, and then back to
the initial point along the other Lagrangian manifold.  This is
discussed in Littlejohn (1990) and in Aquilanti \etal (2007, 2012). In
the case of the Ponzano-Regge formula, the relative phase between the
two branches contained in $\cos(S+\pi/4)$ is $2S$ (dropping the $\pi/4$
which is irrelevant), which therefore is the integral of $\theta$ in
(\ref{theta12j}) along a path starting on $T_1$, running along the
$B$-manifold to $T_2$ and then back along the $A$-manifold to the same
point of $T_1$.  Obviously if the path is traversed in the opposite
direction, we obtain a change of sign in the value of the integral.
These facts are not needed for the proof of the Schl\"afli identity
presented in Sec.~\ref{proof}, but they motivate the general approach.

Such a path can be constructed out of group actions in $G_a$ and
$G_b$.  We start at a point $P \in T_1$, as illustrated in
Fig.~\ref{ABintsect}, which corresponds to a definite tetrahedron in
$\Reals^3$, as illustrated in Fig.~\ref{tetrahedron}.  We write
$\nvec_{123}$, $\nvec_{1'5'6}$, etc for the unit outward pointing
normals to the faces, evaluated at $P$.  For each edge labeled by $r$,
one of the faces meeting at the edge contains $\Jvec_r$ and the other
contains $\Jvec'_r$.  We take the unit normal to the face with
$\Jvec'_r$ and dot it into $\Jvec_r+\Jvec'_r$, and sum over all the
edges.  This gives a Hamiltonian,
	\begin{equation}
	\fl H = \nvec_{1'5'6}\cdot(\Jvec_{11'}+\Jvec_{55'}) +
	    \nvec_{2'6'4}\cdot(\Jvec_{22'}+\Jvec_{66'}) +
	    \nvec_{3'4'5}\cdot(\Jvec_{33'}+\Jvec_{44'}),
	\label{HPtoQ}
	\end{equation}
which is a linear combination of the functions in the $B$-list, and so
generates a motion along the $B$-manifold.  Call the parameter of the
flow generated by this Hamiltonian $\alpha$.  We follow the flow for an
elapsed parameter of $\alpha=\pi$, which rotates each vector $\Jvec_r$,
$\Jvec'_r$ by an angle $\pi$ about an axis orthogonal to itself,
thereby inverting all twelve vectors $\Jvec_r$, $\Jvec'_r$.  This maps
the tetrahedron into its inverted image, taking us along a path such
as $P\to Q$ in Fig.~\ref{ABintsect} to the other branch of the
intersection $T_2$.  At intermediate points along this path we do not
have a tetrahedron or even faces, that is, the four triangle
conditions for the faces are not maintained, but we do have
$\Jvec_r+\Jvec'_r=0$.  At point $Q$ the inverted tetrahedron appears.
Under this flow the twelve spinors are multiplied by elements of
$SU(2)$, for example, $z_1$ and $z'_1$ are transformed according to
	\begin{eqnarray}
	z_1 &\mapsto \exp[-i(\pi/2)\nvec_{1'5'6}\cdot\bsigma] \,z_1,
	\label{z1PQmap} \\
	z'_1 &\mapsto \exp[-i(\pi/2)\nvec_{1'5'6}\cdot\bsigma] \,z'_1.
	\label{zprime1PQmap}
	\end{eqnarray}
In addition, the action integral of the symplectic 1-form $\theta$ in 
(\ref{theta12j}) along the path $P\to Q$ is $\pi H$, the value of the
Hamiltonian along the flow times the elapsed parameter (this applies
to any Hamiltonian that is bilinear in the $z$'s and $z^\dagger$'s).
But since $\Jvec_r+\Jvec'_r=0$ along the $B$-manifold, the Hamiltonian
is zero, and the accumulated action is zero.

Next we follow a path $Q\to P'$ from $T_2$ along the $A$-manifold to a
point on $T_1$, as illustrated in Fig.~\ref{ABintsect}.  The path
follows the Hamiltonian flow generated by
	\begin{equation}
	H=\nvec_{123}\cdot\Jvec_{123} + 
	  \nvec_{1'5'6}\cdot\Jvec_{1'5'6} +
	  \nvec_{2'6'4}\cdot\Jvec_{2'6'4} +
	  \nvec_{3'4'5}\cdot\Jvec_{3'4'5},
	\label{HQtoPprime}
	\end{equation}
that is, $H$ is a sum over all the faces of the normals times the
partial sums of $\Jvec$ vectors in the triangle conditions defining
the faces.  The normals are still evaluated at
$P$ and thus are constant; since these are outward pointing normals at
$P$, they are inward pointing at $Q$.  We follow the flow for an
elapsed parameter of $\alpha=-\pi$, which causes each face to rotate by
an angle $\pi$ about its normal, thereby inverting all vectors
$\Jvec_r$, $\Jvec'_r$ and returning us to the original tetrahedron (at
the point $P'$ in Fig.~\ref{ABintsect}).  At intermediate stages along
the path $Q\to P'$ the diangle conditions are violated, but the faces
triangle conditions continue to hold so the faces are well defined.
On reaching $P'$, the diangle conditions hold once again, and the
tetrahedron is reassembled.  Again, the action integral
vanishes, since the face vectors $\Jvec_{123}$ etc are all zero along
the flow.  As for the action on the spinors, in the case of $z_1$ and
$z'_1$ it is
	\begin{eqnarray}
	z_1 &\mapsto \exp[+i(\pi/2)\nvec_{123}\cdot\bsigma]\,z_1, 
	\label{z1QPprimemap} \\
	z'_1 &\mapsto \exp[+i(\pi/2)\nvec_{1'5'6}\cdot\bsigma]\,z'_1.
	\label{zprime1QPprimemap}
	\end{eqnarray}

The overall effect of these two flows on the primed spinors is the
identity, for example, (\ref{zprime1PQmap}) and
(\ref{zprime1QPprimemap}) imply $z'_1 \mapsto z'_1$, while the unprimed
spinors suffer a change of overall phase, for example, the effect
of (\ref{z1PQmap}) and (\ref{z1QPprimemap}) on $z_1$ is
	\begin{equation}
	z_1 \mapsto \exp[i(\pi/2)(\avec\cdot\bsigma)]
                    \exp[-i(\pi/2)(\bvec\cdot\bsigma)] \, z_1,
	\label{z1PPprimemap}
	\end{equation}
where $\avec=\nvec_{123}$ and $\bvec=\nvec_{1'5'6}$.  But
	\begin{equation}
	\fl\exp[i(\pi/2)(\avec\cdot\bsigma)]
        \exp[-i(\pi/2)(\bvec\cdot\bsigma)] =
	(\avec\cdot\bsigma)(\bvec\cdot\bsigma) =
	\cos\psi_1 + i\sin\psi_1(\jvec_1\cdot\sigma),
	\end{equation}
where $\jvec_1=\Jvec_1/J_1$ and $\psi_1$ is the dihedral
angle at edge 1, as defined previously.  (We have used
$\avec\cdot\bvec=\cos\psi_1$ and $\avec\times\bvec=\sin\psi_1
\jvec_1$.)  But since
	\begin{equation}
	(\jvec_1\cdot\bsigma)z_1 =z_1
	\end{equation}
the overall effect on spinor $z_1$ is 
	\begin{equation}
	z_1 \mapsto e^{i\psi_1}\,z_1,
	\label{z1phasemap}
	\end{equation}
and similarly for the other spinors.  Under the two flows the
tetrahedron in $\Reals^3$ returns to itself but the unprimed spinors
do not, instead they have moved to a different point on their Hopf
circles, which is why the point $P'$ is indicated as different from
the starting point $P$ in Fig.~\ref{ABintsect}.  This motion (along
the Hopf circles) can be regarded as a holonomy or geometric phase
(Berry 1984) in a $U(1)^{12}$ bundle (a Hopf bundle) over angular
momentum space $(\Reals^3)^{12}$.

We close the loop along a path $P'\to P$ as in Fig.~\ref{ABintsect} in
six steps, with each step moving us along the Hopf circle of one of
the spinors $z_r$.  The Hamiltonian for the $r$-th step is $I_r$, with
an elapsed parameter given by $\alpha=2\psi_r$.  For example, in the
first step the Hamiltonian flow of $I_1$ with parameter $2\psi_1$ has
the effect
	\begin{equation}
	z_1 \mapsto e^{-i\psi_1}\, z_1,
	\label{zrPprimeP}
	\end{equation}
thereby cancelling the phase in (\ref{z1phasemap}) and returning $z_1$
to its original value.  The value of $\psi_1$ here is one at the
initial condition $P$, but that is the same value appearing in
(\ref{z1phasemap}), and it is constant on $T_1$ (and therefore it is
the same as at $P'$ and along the $I_1$-flow).  The action along this
$I_1$ flow is $2\psi_1 I_1$.  When all six flows under the $I_r$ are
carried out, the total action is twice the Ponzano-Regge phase
(\ref{Sdef}). 

In summary, the action integrals along the first two legs of the path
$P\to Q\to P' \to P$ vanish, while that along the third gives twice
the Ponzano-Regge phase.  This method of deriving the phase, a part of
the Ponzano-Regge formula, is basically due to Roberts (1999), and it
is much simpler than the derivation we gave in the asymmetrical
$4j$-model of the $6j$-symbol presented in Aquilanti \etal (2012).
The main advantage of the $4j$-model seems to be its close connection
with the spherical phase space of the $6j$-symbol, which leads to an
easy derivation of the amplitude of the Ponzano-Regge formula (due
originally to Wigner (1959)).  The part of this derivation that we
need for this paper is the representation of the Ponzano-Regge phase
as (one half of) the integral of the symplectic 1-form along the path
$P\to Q \to P' \to P$.

\section{Proof of the Schl\"afli Identity}
\label{proof}

The Schl\"afli identity involves variations in the edge lengths $J_r$,
so we must consider families of $A$- and $B$-manifolds and their
intersections in which the values of $I_r$ and $I'_r$ are variable.
To do this we replace the conditions $I_r=I'_r=J_r$ with simply
$I_r=I'_r$, thereby maintaining the equal lengths of vectors $\Jvec_r$
and $\Jvec'_r$.  The $A$-manifold thereby enlarges into a manifold we
call $\Atilde$, defined by $I_r=I'_r$, $r=1,\ldots,6$ plus the four
triangle conditions, which constitute $6+12=18$ independent conditions
and define a 30-dimensional manifold.  Similarly, the $B$-manifold
enlarges to a manifold $\Btilde$, defined by the six diangle
conditions $\Jvec_r+\Jvec'_r=0$.  The conditions $I_r=I'_r$ are not
independent of these and need not be added.  Thus there are
$3\times6=18$ independent conditions, and $\Btilde$ is also
30-dimensional.  Just as the $A$- and $B$-manifolds are Lagrangian,
the $\Atilde$- and $\Btilde$-manifolds are coisotropic, since the
functions defining the level sets have vanishing Poisson brackets
among themselves on the level sets.

As for the 15-dimensional intersection manifolds $T_1$ and $T_2$, we
have been assuming that the values of $J_r$ were such that a
tetrahedron exists with nonzero volume, but when we allow the $J_r$ to
be variable their values will include tetrahedra of zero volume.
These (flat) tetrahedra are equal to their own images under spatial
inversion (modulo a proper rotation), so manifolds $T_1$ and $T_2$
merge at such values of the edge lengths.  Therefore we define a
manifold $\Ttilde=\Atilde\cap\Btilde$; it is connected.  Manifold
$\Ttilde$ is specified by the union of the conditions for $\Atilde$
and $\Btilde$, which are simply the triangle and diangle conditions
taken together (again, $I_r=I'_r$ follows from the diangle
conditions).  These are not independent, because both the triangle and
diangle conditions imply the vanishing of the total angular momentum
(\ref{Jtotdef}); so the number of independent conditions defining
$\Ttilde$ is $4\times3 + 6\times 3 - 3=27$, and the manifold $\Ttilde$
is 21-dimensional (since $48-27=21$).

\begin{figure}[htb]
\begin{center}
\scalebox{0.43}{\includegraphics{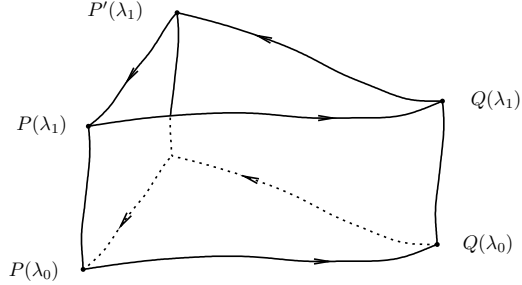}}
\end{center}
\caption[rps]{\label{iron} A family of initial points $P(\lambda)$
with differing values of the $J_r$ generate contours along which the
action integral is twice the Ponzano-Regge phase.  Stokes' theorem is
used to prove the Schl\"afli identity.}
\end{figure}

We now take the point $P$ of Fig.~\ref{ABintsect} and allow it to
sweep out a curve $P(\lambda)$ lying inside $\Ttilde$, for $\lambda_0
\le \lambda \le \lambda_1$, where the original $P$ is now
$P(\lambda_0)$.  For simplicity we assume this curve does not cross
any flat tetrahedra (those with zero volume).  Each $P(\lambda)$
serves as the initial point for a contour with three legs, first
running along the $B$-manifold to a point $Q(\lambda)$, representing
an inverted tetrahedron, then back along the $A$-manifold to a point
$P'(\lambda)$, to a tetrahedron with the original shape, and then
along Hopf circles to return to the original point $P(\lambda)$.  The
values of the $J_r$ are fixed by the initial point $P(\lambda)$, and
are constant along the contour; they are therefore functions of
$\lambda$.  So are the dihedral angles $\psi_r$ and the action
function $S$, which is the integral of $\theta$ along the path.
Starting from points $P(\lambda)$, these paths sweep out a
2-dimensional surface illustrated in Fig.~\ref{iron}, in which the
three segments of the wall, swept out by the three legs of the paths,
are illustrated.
	 
We now integrate the symplectic form $\omega$ along the walls of this
surface.  On the first ($P\to Q$) segment we use $\lambda$ and $\alpha$ as
coordinates, where $\alpha$ is the evolution parameter of the
Hamiltonian (\ref{HPtoQ}).  The integral is
	\begin{equation}
	\int_{\rm seg\,1} \omega =
	\int_{\lambda_0}^{\lambda_1} d\lambda 
	\int_0^\pi d\alpha\;
	\omega\left( \frac{\partial}{\partial\lambda},
	             \frac{\partial}{\partial\alpha}\right).
	\label{leg1face}
	\end{equation}
But $\partial/\partial\alpha$ is the Hamiltonian vector field which we
write as $X_H$, so the integrand can be written
	\begin{equation}
	-(i_{X_H}\omega)\left(\frac{\partial}{\partial\lambda}\right)
	= dH\left(\frac{\partial}{\partial\lambda}\right) = 
	\frac{dH}{d\lambda},
	\label{leg1simplify}
	\end{equation}
where we use Hamilton's equations in the form
	\begin{equation}
	i_{X_H}\omega = -dH,
	\label{Hameqns}
	\end{equation}
and we write the final result as $dH/d\lambda$ (as an ordinary
derivative) because $H$ restricted to the first segment of the wall is
a function only of $\lambda$ (it is independent of $\alpha$).  In
fact, $H$ in (\ref{HPtoQ}) vanishes on the first segment, so
$dH/d\lambda=0$ and the integral of $\omega$ is zero on this segment.  A
similar argument shows that the integral of $\omega$ also vanishes
along the second segment, where the Hamiltonian is (\ref{HQtoPprime}).

Thus the only contribution to the integral of $\omega$ is from the
third segment,
	\begin{equation}
	\int_{\rm walls} \omega =
	\int_{\rm seg\,3} \omega,
	\label{wallint}
	\end{equation}
a result we will use later.  As for the third segment, we break it into
six steps as in Sec.~\ref{PRphase}, where the $r$-th step uses the
Hamiltonian $H=I_r$ and the elapsed parameter $\alpha=2\psi_r$.  For
example, in the first step we have
	\begin{eqnarray}
	\int_{\rm step\,1} \omega &=
	\int_{\lambda_0}^{\lambda_1} d\lambda
	\int_0^{2\psi_1(\lambda)} d\alpha \;
	\omega\left(\frac{\partial}{\partial\lambda},
                    \frac{\partial}{\partial\alpha}\right)
	\nonumber\\ 
	&= \int_{\lambda_0}^{\lambda_1} d\lambda
	\int_0^{2\psi_1(\lambda)} d\alpha \;
	\frac{dI_1}{d\lambda} =
	2\int_{\lambda_0}^{\lambda_1} d\lambda \; \psi_1 
	\frac{dI_1}{d\lambda},
	\label{step1int}
	\end{eqnarray}
where we have transformed the integrand as before.  Doing this for all
six steps (for all $r=1,\ldots,6$), we finally obtain
	\begin{equation}
	\int_{\rm walls} \omega = 2 \sum_r
	\int_{\lambda_0}^{\lambda_1} \psi_r 
	\frac{dI_r}{d\lambda}\, d\lambda.
	\label{omegawalls}
	\end{equation}

Now we apply Stokes' theorem to the integral of $\omega$ over the
walls, which gives
	\begin{equation}
	S(\lambda_1) - S(\lambda_0) = \sum_r 
	\int_{\lambda_0}^{\lambda_1} \psi_r
	\frac{dI_r}{d\lambda} \, d\lambda,
	\label{Stokes}
	\end{equation}
where the loop integrals of $\theta$ along the top and bottom edges of
the walls give the left-hand side and where we have cancelled a
factor of 2.  One can check that the sign conventions of Stokes'
theorem are correct in (\ref{Stokes}).  Finally, differentiating this
with respect to $\lambda_1$ and using (\ref{Sdef}) give the
Schl\"afli identity in the form,
	\begin{equation}
	\sum_r I_r \frac{d\psi_r}{d\lambda} =0.
	\label{Schlafli3}
	\end{equation}
This is our proof of the Schl\"afli identity in Euclidean $\Reals^3$.

\section{The First Reduction}
\label{firstreduction}

This section and the next apply symplectic reduction, the geometrical
method of passing from a symplectic manifold with a symmetry to a
smaller symplectic manifold in which the symmetry has been eliminated.
Symplectic reduction is described in Abraham and Marsden (1978),
Marsden and Ratiu (1999), Cushman and Bates (1997) and Holm (2011).
Basic elements of symplectic reduction that are called upon below are
the symplectic action of a group, the momentum map, and the formation
of the reduced symplectic manifold as the quotient of a level set of
the momentum map by the ``isotropy subgroup'', which in all examples
in this paper is the entire group.

We now apply symplectic reduction to the 48-dimensional phase space
$\Phi_{12j} =(\Phi_{2j})^6$, producing a 36-dimensional phase space
that is almost diffeomorphic to $(T^*SU(2))^6$ (a certain zero level
set must be handled specially, but except for this subset of measure
zero the quotient space is a power of $T^*SU(2)$).  Both manifolds
$\Atilde$ and $\Btilde$ of Sec.~\ref{ABmanifolds} (and hence the $A$-
and $B$- manifolds for all values of the $J_r$) are submanifolds of
the 42-dimensional set given by $I_r-I'_r=0$, $r=1,\ldots,6$.
This is a level set of the momentum map with components $I_r-I'_r$,
whose associated group is $U(1)^6$.  Under the symplectic reduction of
the given (zero) level set of the momentum map by $U(1)^6$ the
Lagrangian $A$- and $B$-manifolds project onto Lagrangian manifolds in
the quotient space, as does the entire construction presented in
Sec.~\ref{proof} and illustrated in Fig.~\ref{iron}.  For example, the
contour for the integral representation of the Ponzano-Regge phase
projects onto a contour in the reduced symplectic manifold giving
another (lower dimensional) integral representation for the same
phase.  Also, the surface swept out when this contour is allowed to
move by varying the $J_r$ projects onto the reduced space, as does the
proof of the Schl\"afli identity based on it.

\subsection{Reduction of $\Phi_{2j}$ by $U(1)$}
\label{reductionPhi2j}

Each $U(1)$ factor of $U(1)^6$ acts on a single factor of
$\Phi_{12j}=(\Phi_{2j})^6$ for a given value of $r$, so to study the
reduction we restrict attention to single value of $r$ and a single
copy of $\Phi_{2j}$.  When we are done we just take the 6-fold product
of the reduction of $\Phi_{2j}$ by $U(1)$ to get that of $\Phi_{12j}$
by $U(1)^6$.  It does not matter which copy of $\Phi_{2j}$ we work
with so in the following we drop the $r$ index.  This same reduction
was discussed by Freidel and Speziale (2010) in connection with the
classical phase space associated with spin networks in loop quantum
gravity.  The following treatment of this reduction differs from that
of Freidel and Speziale in several particulars, some of which will be
pointed out as we proceed.

On the 8-dimensional phase space $\Phi_{2j}=\Complexes^2 \times
\Complexes^2$ the symplectic 1-form and 2-form are
	\begin{equation}
	\theta = iz^\dagger\,dz + iz^{\prime\dagger}\,dz'
	\label{theta2j}
	\end{equation}
and
	\begin{equation}
	\omega=d\theta = idz^\dagger \wedge dz +
	idz^{\prime\dagger} \wedge dz'.
	\label{omega2j}
	\end{equation}
On this space the Hamiltonian function $H=I-I'$ generates the $U(1)$
action,
	\begin{equation}
	z\mapsto e^{-i\alpha/2}z, \qquad
	z' \mapsto e^{+i\alpha/2}z',
	\label{U1action2j}
	\end{equation}
where $\alpha$ is conjugate to $H$.  We are interested in the
7-dimensional level set $L$ given by $I-I'=0$.  This condition can
also be written $z^\dagger z = z^{\prime\dagger}z'$, so $L$ has the
structure of a cone (it is the union of a set of vector spaces passing
through origin $z=z'=0$ of $\Phi_{2j}$).  Except on the subset $I=I'=0$
(the origin, a single point), the orbits of the action
(\ref{U1action2j}) on $L$ are circles of period $4\pi$ in the
coordinate $\alpha$, while the orbit through the origin is just the
origin itself (a single point). 

Because of the change in the dimension of the orbit, the quotient
space $L/U(1)$ is stratified.  Let us define $\Ldot$ as $L$ with the
origin removed,
	\begin{equation}
	\Ldot = L \setminus \{0\},
	\label{Ldotdef}
	\end{equation}
so that on $\Ldot$ the orbits are circles, and $L=\Ldot\cup\{0\}$.  As
for the quotient space, let us define
	\begin{equation}
	\Qdot=\Ldot/U(1),
	\label{Qdotdef}
	\end{equation}
so that as a set the entire quotient space is $Q=\Qdot\cup\{0\}$.  In
the following we will concentrate on $\Ldot$ and $\Qdot$, the latter
of which is a symplectic manifold of dimension 6.  Our aim will be to
identify $\Qdot$ topologically, to find convenient coordinates on it,
and to find its symplectic form.

It turns out that $\Qdot$ is symplectomorphic to $T^*SU(2)$ with a
certain subset removed.  Freidel and Speziale (2010) have stated that
the quotient space is diffeomorphic to $T^*SU(2)$, but this is not
quite right, since a subset of measure zero has to be handled
separately.  Stratifications of this sort are treated in the theory of
``symplectic implosion'' (Guillemin, Jeffrey and Sjamaar 2002).

It is somewhat awkward to demonstrate the relation between $\Qdot$ and
$T^*SU(2)$ in the coordinates and in terms of the symplectic structure
given so far.  Instead it is convenient to use a symplectic map which
we call $K_2$, taking us from $\Phi_{2j}$ to another symplectic
manifold which we call $\Phi_{2j}^*$, and to carry out the symplectic
reduction on $\Phi_{2j}^*$.  When this is done, we can use $K_2$ to
pull the reduction back to $\Phi_{2j}$.  As a manifold, $\Phi_{2j}^*$
is $\Complexes^2 \times \Complexes^2$ with coordinates $(z,z')$, just
as with $\Phi_{2j}$, but the symplectic forms on $\Phi_{2j}^*$ are
	\begin{equation}
	\theta=iz^\dagger \, dz - iz^{\prime\dagger}\,dz',
	\label{theta2jstar}
	\end{equation}
and
	\begin{equation}
	\omega=d\theta=idz^\dagger \wedge dz 
	-idz^{\prime\dagger}\wedge dz',
	\label{omega2jstar}
	\end{equation}
that is, with a change in sign in the primed term relative to the
symplectic forms (\ref{theta2j}) and (\ref{omega2j}) on $\Phi_{2j}$.
(The notation $\Phi_{2j}^*$ is slightly illogical, since the sign is
changed only in the primed term of the symplectic form.)  Functions
$I$, $I'$, $\Jvec$ and $\Jvec'$ are defined on $\Phi_{2j}^*$ by the
same formulas (\ref{IJdef}) as on $\Phi_{2j}$, but because of the
change in sign in the symplectic structure we have the Poisson bracket
relations $\{J'_i,J'_j\} = -\epsilon_{ijk}\,J'_k$ on $\Phi_{2j}^*$.
The other Poisson bracket relations on $\Phi_{2j}^*$ are the same as
on $\Phi_{2j}$, namely, $\{I,J_i\}=\{I',J'_i\}=0$ and
$\{J_i,J_j\}=\epsilon_{ijk}\,J_k$.  Also, any function of
$z,z^\dagger$ Poisson commutes with any function of
$z',z^{\prime\dagger}$.

We now make a digression to explain some maps that will be used to
construct the map $K_2:\Phi_{2j}\to\Phi_{2j}^*$.

\subsection{Some Useful Maps}

In this section we consider the phase spaces $\Phi=(\Complexes^2,
idz^\dagger\wedge dz)$ and $\Phi^*=(\Complexes^2, -idz^\dagger\wedge
dz)$.  The 2-component spinor $z$ is a coordinate on both $\Phi$ and
$\Phi^*$, but the symplectic forms are of opposite sign. Space $\Phi$
is the same one introduced in Sec.~\ref{phasespaces}.  Notice that the
phase spaces discussed in Sec.~\ref{reductionPhi2j} can be written
$\Phi_{2j} = \Phi\times\Phi$ and $\Phi_{2j}^* = \Phi\times\Phi^*$,
where it is understood that symplectic forms add under the Cartesian
product.  

For motivational reasons we explain the semiclassical significance of
$\Phi$ and $\Phi^*$. Space $\Phi$ is the phase space corresponding to
the Hilbert space $\Hspace=L^2(\Reals^2)$. We view $\Hspace$ as the space
of wave functions of the quantum 2-dimensional harmonic oscillator,
which is used in the Schwinger (1952) and Bargmann (1962) treatment of
the representations of $SU(2)$.  We also view $\Hspace$ as a space of
Dirac ket vectors.  To say that $\Hspace$ corresponds to $\Phi$ means
at least two things.  One is that linear operators from $\Hspace$ to
itself are mapped into functions on $\Phi$ by means of the Weyl
correspondence (Berry 1977, Balasz and Jennings 1984, Ozorio de
Almeida 1998).  Another is that wave functions $\psi(x_1,x_2)$ in
$\Hspace$ are represented by Lagrangian manifolds in $\Phi$; the phase
of the semiclassical approximation to the wave function is the
integral of the symplectic 1-form $p\,dx$ (which differs from
$iz^\dagger\, dz$ by an exact differential) along the Lagrangian
manifold, times $1/\hbar$.

We define $\Hspace^*$ as the space dual to $\Hspace$, that is,
$\Hspace^*$ contains bra vectors or complex valued linear functionals
on $\Hspace$.  We can also think of $\Hspace^*$ as containing wave
functions $\psi^*(x_1,x_2)$, that is, a wave function in $\Hspace^*$
is the image under the metric $\Ghat:\Hspace\to\Hspace^*:
\psi(x_1,x_2)\mapsto\psi^*(x_1,x_2)$ of some wave function in $\Hspace$.
The metric $\Ghat$ is an antilinear map.  If $\psi(x_1,x_2)\in\Hspace$
corresponds to a Lagrangian manifold in $\Phi$, then
$\psi^*(x_1,x_2)\in\Hspace^*$ corresponds to the same Lagrangian
manifold in $\Phi^*$.  However, since the symplectic 1-form in $\Phi^*$
is the opposite of that in $\Phi$, its integral along the Lagrangian
manifold in $\Phi^*$ is opposite the corresponding integral in $\Phi$.
This inverts the phase of the semiclassical approximation to the wave
function, which is just what we want for the complex conjugated wave
function.

These considerations motivate the following definition of the
classical map $G$ corresponding to the metric $\Ghat$:
	\begin{equation}
	G:\Phi\to\Phi^*:z \mapsto z.
	\label{Gdef}
	\end{equation}
In other words, $G$ is the identity as far as the coordinates $z$ are
concerned, but as a map it connects different spaces.  Moreover, it is
antisymplectic, since the push-forward of the symplectic form on
$\Phi$ under $G$ is minus the symplectic form on $\Phi^*$.

Another map between $\Phi$ and $\Phi^*$ is defined by
	\begin{equation}
	K:\Phi\to\Phi^*:z \mapsto U_0 \zbar,
	\label{Kdef}
	\end{equation}
where $\zbar$ is the complex conjugated column spinor and where
	\begin{equation}
	U_0 = e^{-i(\pi/2)\sigma_y} =
	\left(\begin{array}{cc}
	0 & -1 \\
	1 & 0
	\end{array}\right).
	\label{U0def}
	\end{equation}
Under $K$ the symplectic form $idz^\dagger\wedge dz$ on $\Phi$ is
pushed forward into the form $-idz^\dagger\wedge dz$ on $\Phi^*$ which
is the same as the symplectic form on $\Phi^*$ so the map $K$ is
symplectic.  It is the classical analog of the quantum map $K$
discussed in Aquilanti \etal (2012), modulo phase conventions that we
will not go into here.  Suffice it to say that the quantum map $K$ is
a linear map between $\Hspace$ and $\Hspace^*$, unlike $\Ghat$, which
is antilinear.  We see that linear maps in quantum mechanics
correspond to symplectic maps classically, and quantum antilinear maps
to classical antisymplectic maps.

Under $K$ the functions $J_i$ on $\Phi$ are pushed forward into the
functions $-J_i$ on $\Phi^*$, while $I$ becomes $I$.  Since $K$ is
symplectic the formation of the Poisson bracket commutes with the
push-forward; for example, computing $\{J_1,J_2\}$ on $\Phi$ gives
$J_3$, which pushes forward into $-J_3$; but $J_1$ and $J_2$ push
forward into $-J_1$ and $-J_2$ whose Poisson bracket on $\Phi^*$ is
computed according to $\{-J_1,-J_2\} = \{J_1,J_2\} = -J_3$, where the
final minus sign comes from the inverted symplectic structure on
$\Phi^*$.  The answers are the same.

A third map is time-reversal, defined by
	\begin{equation}
	\Theta:\Phi\to\Phi: z \mapsto U_0\zbar.
	\label{Thetadef}
	\end{equation}
In the coordinates $z$, $\Theta$ is defined by the same equation as
$K$, but unlike $K$ it maps $\Phi$ to $\Phi$ so it is antisymplectic.
We call $\Theta$ time-reversal because it is the classical analog of
the quantum time-reversal operator (Messiah 1966), an antilinear
operator that inverts the direction of angular momenta.  The action
shown in (\ref{Thetadef}) is the same as that of the quantum time-reversal
operator on the spinor of a spin-$1/2$ particle.  Under $\Theta$, the
functions $J_i$ go into $-J_i$, while $I$ goes into $I$, just as under
$K$; but now these function are on $\Phi$, not $\Phi^*$.  We note that
the three maps introduced are related by
        \begin{equation}
          \Theta=G^{-1} \circ K.  
          \label{ThetaGK}
        \end{equation}

Considered as a map from $\Complexes^2$ to $\Complexes^2$,
time-reversal is an antiunitary operator satisfying
$\Theta^\dagger\Theta=1$. It has the property
	\begin{equation}
	\Theta^\dagger \sigma_i \Theta = -\sigma_i.
	\label{Thetasigma}
	\end{equation}
From this follows the cited transformation properties of $I$ and
$\Jvec$.  For example, using an obvious notation for the scalar
product of two 2-component spinors and the properties of antilinear
operators we have
	\begin{equation}
	\fl I=\frac{1}{2}z^\dagger z = \frac{1}{2} \langle z,z\rangle \mapsto
	\frac{1}{2}\langle\Theta z,\Theta z\rangle =
	\frac{1}{2}\overline{\langle z,\Theta^\dagger\Theta z\rangle}
	=\frac{1}{2}\overline{\langle z,z\rangle} = \bar{I} = I,
	\label{IunderTheta}
	\end{equation}
where we have used the fact that $I$ is real.  Similarly, we have
	\begin{eqnarray}
	J_i &=\frac{1}{2}z^\dagger \sigma_i z = 
	\frac{1}{2}\langle z,\sigma_i z\rangle \mapsto 
	\frac{1}{2}\langle\Theta z,\sigma_i \Theta z\rangle 
	\nonumber \\ &=
	\frac{1}{2}\overline{\langle z,\Theta^\dagger 
          \sigma_i\Theta z\rangle} =
	-\frac{1}{2}\overline{\langle z,\sigma_i z\rangle} 
        = -\bar{J_i} = -J_i,
	\label{JunderTheta}
	\end{eqnarray}
since $J_i$ is real.  Equation~(\ref{Thetasigma}) also implies that
time-reversal commutes with rotations, that is,
	\begin{equation}
	\Theta^\dagger g \Theta = g,
	\label{Thetarotations}
	\end{equation}
for all $g\in SU(2)$.  This follows if we write $g$ in the axis-angle
form (\ref{uaxisangle}) since the antilinear $\Theta$ inverts the
signs of both $\sigma_i$ and $i$.

\subsection{Reduction of $\Phi_{2j}^*$ by $U(1)$}
\label{reductionPhi2jstar}  

We return now to the discussion interrupted at the end of
Sec.~\ref{reductionPhi2j}.  We define the symplectic map
	\begin{equation}
	K_2:\Phi_{2j} \to \Phi_{2j}^*:(z,z') \mapsto
	(z,Kz'),
	\label{K2def}
	\end{equation}
in which the $K$ map is applied only to the second argument.  The 2
subscript on $K_2$ is a reminder of this fact.  Note that $\Phi_{2j}$
may be regarded as the phase space corresponding to vectors
$\ket{\psi}\ket{\phi}$, where $\ket{\psi}$ and $\ket{\phi}$ are state
vectors for the 2-dimensional harmonic oscillator, while $\Phi_{2j}^*$
may be regarded as the phase space corresponding to vectors
$\ketbra{\psi}{\phi}$.

We will carry out the reduction of $\Phi_{2j}$ by $U(1)$ described in
Sec.~\ref{reductionPhi2j} by mapping the entire construction over
to $\Phi_{2j}^*$ via $K_2$.  Under the symplectic map $K_2$ symplectic
group actions, momentum maps and level sets in $\Phi_{2j}$ are mapped
into the same types of objects in $\Phi_{2j}^*$.  For example, the
function $I-I'$ on $\Phi_{2j}$ maps into $I-I'$ on $\Phi_{2j}^*$,
whose Hamiltonian flow on $\Phi_{2j}^*$ is given by
	\begin{equation}
	(z,z') \mapsto (e^{-i\alpha/2}z,
	e^{-i\alpha/2} z'),
	\label{U1action2jstar}
	\end{equation}
with a change in sign on the second half as compared with
(\ref{U1action2j}) due to the change in sign in the second term of the
symplectic form.  This is the $U(1)$ action on $\Phi_{2j}^*$ by which
we carry out the reduction.  The fact that both spinors $z$ and $z'$
transform in the same way under this action is the main fact that
makes the reduction on $\Phi_{2j}^*$ more transparent than that on
$\Phi_{2j}$.  

The level set $I-I'=0$ is the same 7-dimensional cone discussed in
Sec.~\ref{reductionPhi2j}.  We use the same symbol $L$ for it on
$\Phi_{2j}^*$ as on $\Phi_{2j}$, and we define $\Ldot$ and $\Qdot$
exactly as in (\ref{Ldotdef}) and (\ref{Qdotdef}).  Obvious functions
on $\Phi_{2j}^*$ that are constant on the orbits
(\ref{U1action2jstar}) and which therefore can be projected onto the
quotient space are quantities bilinear in $z^\dagger$ and $z$ or in
$z^{\prime\dagger}$ and $z'$, including $I$, $I'$, $\Jvec$ and
$\Jvec'$.

Another set of functions that can be projected onto the quotient space
is constructed as follows.  We begin by defining a map from
$\Complexes^2$ to $2\times 2$ complex matrices.  If $z\in
\Complexes^2$, then the map is
	\begin{equation}
	z \mapsto M(z) = (z,\Theta z) =
	\left(\begin{array}{cc}
	z_1 & -\zbar_2 \\
	z_2 & \zbar_1 \end{array}\right),
	\label{ztomatrixdef}
	\end{equation}
where $(z,\Theta z)$ means the matrix whose two columns are $z$ and
$\Theta z$.  We note that $\det M(z)=2I$, so if $I\ne0$ then $M(z)$
has an inverse, and, in fact, 
	\begin{equation}
	M(z)/\sqrt{2I} \in SU(2).
	\label{MvsSU2}
	\end{equation}
Thus $M(z)$ can be identified with a quaternion, that is, it can be
written $q_0-i\qvec\cdot\bsigma$, where $(q_0,\qvec)$ is a real
4-vector.

Now let $(z,z') \in \Ldot$ (so that $I=I'\ne0$), and let $g$ be an
element of $SU(2)$ such that 
	\begin{equation}
	z=gz'.
	\label{gdef}
	\end{equation}
Then by (\ref{Thetarotations}) we have $\Theta z = g\Theta z'$, and
	\begin{equation}
	M(z) = g M(z'),
	\label{Mgeqn}
	\end{equation}
or,
	\begin{equation}
	g = M(z)M(z')^{-1} = \frac{1}{2I} 
	\left(\begin{array}{cc}
	z_1 & -\zbar_2 \\
	z_2 & \zbar_1\end{array}\right) 
	\left(\begin{array}{cc}
	\zbar'_1 & \zbar'_2 \\
	-z'_2 & z'_1\end{array}\right).
	\label{gsolve}
	\end{equation}
Thus, for $(z,z')\in \Ldot$, there is a unique $g\in SU(2)$ such that
$z=gz'$, given as an explicit function of $z$ and $z'$ by
(\ref{gsolve}).  Moreover, the matrix $g$, that is, its components,
are constant along the orbits (\ref{U1action2jstar}).  This is obvious
since $g$ in $z=gz'$ remains the same if both $z$ and $z'$ are
multiplied by the same phase.

We remark that the map $z\to M(z)$ can be used to show the equivalence
of the Hopf circles as defined by us, that is, $z\mapsto e^{-i\alpha/2}z$,
following the Hamiltonian flow of $I$ on
$\Phi=(\Complexes^2,idz^\dagger \wedge dz)$, with the definition used
by Freidel and Speziale (2010) and other authors, which is
	\begin{eqnarray}
	M(z) &\mapsto M(z) u(\zhat,\alpha) =
	(z,\Theta z) \left(\begin{array}{cc}
	e^{i\alpha/2} & 0 \\
	0 & e^{-i\alpha/2}\end{array}\right) \nonumber\\
	&=
	(e^{i\alpha/2}z, \Theta e^{i\alpha/2}z) =
	M(e^{i\alpha/2}z),
	\label{otherHopfcircle}
	\end{eqnarray}
where $u(\zhat,\alpha)$ is a spin-$1/2$ rotation of angle $\alpha$
about the $z$-axis and where we use the antilinearity of $\Theta$.  In
the latter definition the Hopf circles are identified with the cosets
in $SU(2)$ with respect to the subgroup $U(1)$ of rotations about the
$z$-axis.  The two definitions of the Hopf circles are the same, apart
from a switch in the direction of traversal.  

We use these results to construct global coordinates on $\Ldot$.  A
point of $\Ldot$ can be identified with a pair $(z,z')$ such that
$I=I'\ne0$.  But in view of the uniqueness of $g\in SU(2)$ such that
$z=gz'$, such a point can also be uniquely identified by the pair
$(z,g)$ or the pair $(g,z')$, where $z=gz'$ and where $|z|\ne0$ and
$|z'|\ne0$.  Moreover, any such pair, in either version, corresponds
to a unique point of $\Ldot$, and either pair can be taken as
coordinates on $\Ldot$.  Defining
        \begin{equation}
          \Complexesdot^2 = \Complexes^2 \setminus \{0\},
          \label{C2dotdef}
        \end{equation}
we see that $\Ldot$ is diffeomorphic to $\Complexesdot^2\times SU(2)$
and to $SU(2)\times\Complexesdot^2$  via the two coordinate systems
$(z,g)$ and $(g,z')$.

When we subject a point of $\Ldot$ with coordinates $(z,g)$ to the
action (\ref{U1action2jstar}), we have $z\mapsto e^{-i\alpha/2}z$,
$g\mapsto g$; or, in coordinates $(g,z')$, we have $z'\mapsto
e^{-i\alpha/2}z'$, $g \mapsto g$.  In either case, the action is a
motion along the Hopf circle in one of the two ($z$ or $z'$) factors,
while $g$ is left invariant.  So the quotient space is obtained by
dividing the $\Complexesdot^2$ factor in the two factorizations of
$\Ldot$ by the Hopf action, while leaving the $SU(2)$ factor alone.
The result is
       \begin{equation}
         \Qdot\cong \Realsdot^3\times SU(2) \cong
         SU(2) \times \Realsdot^3,
         \label{quotspace}
       \end{equation}
where $\cong$ means ``is diffeomorphic to'' and where
       \begin{equation}
         \Realsdot^3 = \Reals^3 \setminus \{0\} 
         =\frac{\Complexesdot^3}{U(1)}.
         \label{R3dotdef}
       \end{equation}
In the final quotient operation the projection is the Hopf map $\pi_H$
discussed below (\ref{IJdef}), so that coordinates on $\Qdot$ are
either $(\Jvec,g)$ or $(g,\Jvec')$, where $J_i = (1/2)z^\dagger
\sigma_i z$ and $J'_i = (1/2)z^{\prime\dagger}\sigma_i z'$.  Vectors
$\Jvec$ and $\Jvec'$ belong to $\Realsdot^3$ and are related by
       \begin{equation}
         \Jvec=R(g)\Jvec',
         \label{adjointeqn}
         \end{equation}
where $R(g)$ is defined by (\ref{Rijdef}).  This can be proved with
the relation,
       \begin{equation}
         g^\dagger \sigma_i g = \sum_j R_{ij}(g) \,\sigma_j,
         \label{SU2adjoint}
       \end{equation}
essentially a statement of the coadjoint representation of $SU(2)$.

The cotangent bundle $T^*SU(2)$ is diffeomorphic to $\Reals^3 \times
SU(2)$ or $SU(2) \times \Reals^3$, where $\Reals^3$ is the dual of the
Lie algebra $\Liealgebra{g}^*$.  This is proved by taking an element
of $T^*SU(2)$, which is a 1-form at a point $g\in SU(2)$, and pulling
it back to the identity by either left or right translations.  The
pulled-back form is then expressed in some basis in
$\Liealgebra{g}^*$.  In view of (\ref{quotspace}) our space $\Qdot$ is
diffeomorphic to $T^*SU(2)$ minus the zero section, which we write as
         \begin{equation}
           \Qdot \cong \TstarSU2dot = 
             \{ \alpha\in T^*SU(2)\, \vert\, \alpha\ne0\}.
           \label{TstarSU2dotdef}
         \end{equation}
Having found the topology of our quotient space and convenient
coordinates on it, we now work out the symplectic form and show that
it is identical to the natural symplectic form on $\TstarSU2dot$.

\subsection{Symplectic Structure on $\Qdot$}
\label{symplecticstrucquot}

The symplectic structure on $\Qdot$ can be obtained simply by working out
the Poisson brackets of the coordinates $(\Jvec,g)$ or $(g,\Jvec')$ on
$\Qdot$, where $g$ means the matrix with components $g_{\mu\nu}$ (a
matrix in $SU(2)$, not to be confused with a metric).  These Poisson
brackets can be computed on $\Phi_{2j}^*$ using the symplectic
structure (\ref{omega2jstar}), and they survive unaltered when
projected onto $\Qdot$.  The brackets involving $\Jvec$ and $\Jvec'$
are not difficult, since these vectors are the generators of $SU(2)$
actions on $z$ and $z'$, and the brackets
$\{g_{\mu\nu},g_{\alpha\beta}\}$ can be worked out from the explicit
expressions (\ref{gsolve}).  But the latter calculation is lengthy and
uninspiring.  A better approach is to work with the symplectic form.

The symplectic 2-form (\ref{omega2jstar}) can be restricted to
$\Ldot$, where it is no longer symplectic but it is the pull-back
under the projection map of the symplectic form on $\Qdot$.  This can
be used to find the symplectic form on $\Qdot$.  As a practical
matter, this means using the defining relation of $L$ (that is,
$I-I'=0$) to restrict $\omega$ to $\Ldot$, and then expressing the
result in terms of the coordinates on $\Qdot$ and their differentials.
Since we have two coordinate systems on $\Qdot$, this is to be done in
two different ways.

Actually this process can be carried out on the symplectic 1-form
(\ref{theta2jstar}).  Since $I-I'=0$ implies $z=gz'$ for $g\in SU(2)$,
on $\Ldot$ we can set $z'=g^\dagger z$ and $dz'=dg^\dagger z+
g^\dagger \, dz$, or,
	\begin{eqnarray}
	\theta&=iz^\dagger \, dz - iz^{\prime\dagger}\,dz'
	=-i z^\dagger g\,dg^\dagger z
	=-i \tr((zz^\dagger)g\,dg^\dagger) \nonumber\\
	&=-iI\tr(g\,dg^\dagger) -i \Jvec\cdot\tr(\bsigma g\,dg^\dagger),
	\label{thetaLdot}
	\end{eqnarray}
where we have used
	\begin{equation}
	zz^\dagger = I + \Jvec\cdot\bsigma.
	\label{zzdaggereqn}
	\end{equation}
In (\ref{zzdaggereqn}) $I$ is the function defined in (\ref{IJdef}),
not the identity matrix; multiplication of $I$ by the identity matrix
is understood.  However, since $g\in SU(2)$ we have $gg^\dagger =
g^\dagger g=1$ and $\det g=1$, or,
	\begin{equation}
	\eqalign{
	dg \, g^\dagger + g \,dg^\dagger &=0, \\
	dg^\dagger g + g^\dagger \, dg &=0,}
	\label{gunitary}	
	\end{equation}
and
	\begin{equation}
	\tr (g^\dagger \, dg)= \tr (g\, dg^\dagger)=0.
	\label{detg1}
	\end{equation}
These imply
	\begin{equation}
	\theta=-i\Jvec\cdot\tr(\bsigma g\,dg^\dagger) =
	i\Jvec\cdot\tr(\bsigma dg\,g^\dagger).
	\label{thetaQdot}
	\end{equation}
Similarly eliminating $z$ in favor of $g$ and $z'$ gives
	\begin{equation}
	\theta=i\Jvec'\cdot\tr(\bsigma g^\dagger dg) =
	-i\Jvec'\cdot\tr(\bsigma\,dg^\dagger g).
	\label{thetaQdotprime}
	\end{equation}
These are four expressions for the symplectic 1-form on $\Qdot$. 

\subsection{Symplectic Structure on $T^*SU(2)$}
\label{symplecticstrucTstarSU2}

In the following we regard $SU(2)$ as a 3-dimensional submanifold of
$2\times2$ complex matrix space, that is, $\Complexes^4$.  At some
risk of confusion we use $g$ to denote a point of $SU(2)$ as well as a
complex matrix with components $g_{\mu\nu}$.  At the identity element
the tangent space (the Lie algebra) is spanned by three tangent
vectors $\{e_i,i=1,2,3\}$, defined by
	\begin{equation}
	e_i \,g = -\frac{i}{2} \sigma_i, \qquad
	e_i \,g^\dagger = \frac{i}{2}\sigma_i.
	\label{eidef}
	\end{equation}
In these equations $g$ can be thought of as a matrix of complex
functions (the components $g_{\mu\nu}$) defined on matrix space or on
the $SU(2)$ submanifold thereof; and $e_i$ can be thought of either as
vectors in $T^*SU(2)$ at the identity or vectors in the tangent space
to matrix space at the same point.  In the following we prefer to
write these equations in a slightly different form,
	\begin{equation}
	e_i (dg) = -\frac{i}{2} \sigma_i, \qquad
	e_i (dg^\dagger) = \frac{i}{2}\sigma_i.
	\label{eidefdg}
	\end{equation}
where now $dg$ means the matrix of differential forms, the
differentials of the functions just introduced, either on matrix space
or restricted to the $SU(2)$ submanifold, and where the vector $e_i$
acts on a differential form in the usual way.  These equations define
$e_i$, which can also be written 
	\begin{equation}
	e_i = -\frac{i}{2}\sum_{\mu\nu} (\sigma_i)_{\mu\nu}
	\frac{\partial}{\partial g_{\mu\nu}}
	+\frac{i}{2}\sum_{\mu\nu} (\sigma_i)_{\mu\nu}
	\frac{\partial}{\partial {\bar g}_{\nu\mu}},
	\label{eiasdiffops}
	\end{equation}
expressing them as linear combinations of tangent vectors to matrix
space.  These linear combinations are tangent to the submanifold
$SU(2)$ at the identity.

Now transporting $e_i$ to an arbitrary point $g\in SU(2)$ by left and
right translations, we obtain the left- and right-invariant vector
fields on $SU(2)$, defined by their actions on the matrices of
differential forms $dg$ and $dg^\dagger$,
	\begin{equation}
	\eqalign{
	X^L_i(dg) &= -\frac{i}{2}g\sigma_i, \qquad
	X^L_i(dg^\dagger) = \frac{i}{2}\sigma_i g^\dagger,\\
	X^R_i(dg) &= -\frac{i}{2}\sigma_i g, \qquad
	X^R_i(dg^\dagger) = \frac{i}{2} g^\dagger \sigma_i,}
	\label{XLXRdef}
	\end{equation}
where the vector fields are evaluated at $g\in SU(2)$ (the same $g$
that appears on the right-hand sides).  These definitions imply
	\begin{equation}
	[X^L_i,X^L_j]=\epsilon_{ijk}\, X^L_k, \qquad
	[X^R_i,X^R_j]=-\epsilon_{ijk} \, X^R_k,
	\label{Xcommrels}
	\end{equation}
and
	\begin{equation}
	[e_i,e_j] = \epsilon_{ijk} \, e_k,
	\label{Liealgbra}
	\end{equation}
showing that the Lie algebra in the differential geometric sense agrees
with the matrix Lie algebra of the matrices $\{-(i/2)\sigma_i,
i=1,2,3\}$.  

Now we define 1-forms on $SU(2)$,
	\begin{equation}
	\eqalign{
	\rho^i_L &= i\tr(\sigma_i g^\dagger\,dg) =
	-i\tr(\sigma_i \,dg^\dagger g), \\
	\rho^i_R &= -i\tr(\sigma_i g\,dg^\dagger) =
	i\tr(\sigma_i\,dg\,dg^\dagger).}
	\label{rhoLRdef}
	\end{equation}
These could also be thought of as 1-forms on matrix space, but the two
different versions given are only equal when restricted to $SU(2)$.
Then by a direct calculation we find
	\begin{equation}
	X^L_i(\rho^j_L) = i X^L_i[\tr(\sigma_j g^\dagger\,dg)]
	=\frac{1}{2}\tr(\sigma_j g^\dagger g\sigma_i)
	=\delta_{ij},
	\label{XrhoLON}
	\end{equation}
and similarly we find $X^R_i(\rho_R^j) = \delta_{ij}$.  Therefore
$\rho_L^i$ and $\rho_R^i$ are respectively the left- and
right-invariant 1-forms on $SU(2)$.  Then (\ref{Xcommrels}) implies
	\begin{equation}
	d\rho^i_L = -\frac{1}{2} \epsilon_{ijk}\, 
	\rho^j_L \wedge \rho^k_L, \qquad
	d\rho^i_R = \frac{1}{2} \epsilon_{ijk}\,
	\rho^j_R \wedge \rho^k_R.
	\label{drhorels}
	\end{equation}

The definitions (\ref{rhoLRdef}) were motivated by the expressions
(\ref{thetaQdot}) and (\ref{thetaQdotprime}), but note that $\rho_L^i$
and $\rho_R^i$ are forms on $SU(2)$, while the forms in
(\ref{thetaQdot}) and (\ref{thetaQdotprime}) are on $\Qdot$.  

We are also interested in the symplectic 1-form on $T^*SU(2)$, which
is constructed according to the procedure described by Arnold (1989),
p.~202.  On any manifold $M$ we let $\alpha\in T^*M$, so that $\alpha$
is a 1-form at some point $x\in M$.  Then we define $\theta$, the
symplectic 1-form on $T^*M$, by $\theta|_\alpha = \pi^* \alpha$, where
$\pi:T^*M\to M$ is the bundle projection.  To apply this procedure to
$T^*SU(2)$ we let $\alpha \in T^*SU(2)$ so that $\alpha$ is a 1-form
on $SU(2)$ at a point $g\in SU(2)$, and we express $\alpha$ in terms
of its components with respect to the bases $\{\rho_L^i\}$ and
$\{\rho_R^i\}$ at $g$,
	\begin{equation}
	\alpha = \sum_i \alpha^L_i \rho^i_L 
	=\sum_i \alpha^R_i \rho^i_R.
	\label{alphacomponentsdef}
	\end{equation}
Then $(\alpha^R_i,g)$ and $(g,\alpha^L_i)$ are two coordinate
systems on $T^*SU(2)$.  In terms of these coordinates the symplectic
1-form is
	\begin{equation}
	\theta = \sum_i \alpha^L_i \rho^i_L 
	=\sum_i \alpha^R_i \rho^i_R,
	\label{thetaTstarSU2}
	\end{equation}
that is, the same formula but now with a reinterpretation of
$\rho^i_L$ and $\rho^i_R$, which have been pulled back from
$SU(2)$ to become forms on $T^*SU(2)$.

\subsection{Identifying $\Qdot$ and $\TstarSU2dot$}
\label{identifying}

We now have two coordinate systems on $\Qdot$, $(J_i,g)$ and
$(g,J'_i)$; and two on $\TstarSU2dot$, $(\alpha^R_i,g)$ and
$(g,\alpha^L_i)$.  Comparing the forms (\ref{thetaQdot}) and
(\ref{thetaTstarSU2}) and using the definitions (\ref{rhoLRdef}), we
see that if coordinates $(\alpha^R_i,g)$ on $T^*SU(2)$ are identified
with $(J_i,g)$ on $\Qdot$ then the symplectic 1-forms agree.  That is,
if we just set $J_i=\alpha^R_i$, then the identification between
$\Qdot$ and $\TstarSU2dot$ is a symplectomorphism.  Similarly, setting
$J'_i=\alpha^L_i$ produces another symplectomorphism between $\Qdot$
and $\TstarSU2dot$.  These are the same symplectomorphisms, because on
$T^*SU(2)$ the left- and right- components of a form are related by
the coadjoint representation,
	\begin{equation}
	\alpha^R_i = \sum_j R_{ij}(g) \,\alpha^L_j,
	\label{coadjointSU2}
	\end{equation}
where both forms are evaluated at the group element $g$.  But under our
coordinate transformations this is the same as (\ref{adjointeqn}) on
$\Qdot$.  With these identifications, we can write the symplectic
1-form on $\Qdot \cong \TstarSU2dot$ as
	\begin{equation}
	\theta = \sum_i J_i\,\rho_R^i = \sum_i J'_i \rho_L^i.
	\label{thetaQdot1}
	\end{equation}
This is valid on the version of $\Qdot\cong\TstarSU2dot$ that is
derived from $\Phi_{2j}^*$ by symplectic reduction.  There is another
version of $\Qdot$ that we now describe.

\subsection{Pulling the Reduction Back to $\Phi_{2j}$}
\label{pullingbackreduction}

We now pull back the reduction that we have just carried out on
$\Phi_{2j}^*$ to $\Phi_{2j}$ via the symplectic map $K_2$, defined in
(\ref{K2def}).  The relation defining the coordinates $(z,g)$ or
$(g,z')$ on $\Ldot$, which is $z=gz'$ on $\Phi_{2j}^*$, becomes
$z=g\Theta z'$ on $\Phi_{2j}$.  When this is used to express the
symplectic form (\ref{theta2j}) on $\Ldot$ and thence on $\Qdot$, we
obtain
	\begin{eqnarray}
	\theta &=-i\Jvec\cdot\tr(\bsigma g\,dg^\dagger) =
	i\Jvec\cdot\tr(\bsigma dg\,g^\dagger)
	\nonumber\\
	&=-i\Jvec'\cdot\tr(\bsigma g^\dagger dg) =
	i\Jvec'\cdot\tr(\bsigma\,dg^\dagger g),
	\label{thetaQdot2}
	\end{eqnarray}
that is, with a minus sign in the primed expressions relative to
(\ref{thetaQdotprime}).  This sign is easily understood as the effect of
$K_2$ on the functions $\Jvec'$ (that is, they are mapped into
$-\Jvec'$).  Equation~(\ref{thetaQdot1}) in turn implies that when
reducing from $\Phi_{2j}$, $\Qdot$ is identified with $\TstarSU2dot$
by $J_i=\alpha_R^i$ and $J'_i = -\alpha_L^i$ (with a minus sign in the
primed expression).  Thus the symplectic 1-form on $\TstarSU2dot$, when
obtained by reduction from $\Phi_{2j}$, is
	\begin{equation}
	\theta= \sum_i J_i\, \rho_R^i = -\sum_i J'_i \,\rho_L^i.
	\label{thetaQdot3}
	\end{equation}
Similarly, the relation (\ref{adjointeqn}), which applies to
$\Qdot\cong\TstarSU2dot$ when derived from $\Phi_{2j}^*$, becomes 
        \begin{equation}
        \Jvec=-R(g)\Jvec',
        \label{adjointeqn1}
        \end{equation}
when $\TstarSU2dot$ is obtained from $\Phi_{2j}$.  

From (\ref{thetaQdot3}) we obtain the symplectic 2-form $\omega=d\theta$,
	\begin{eqnarray}
	\omega&=\sum_i dJ_i\wedge \rho_R^i +
	\frac{1}{2}\sum_{ijk}\epsilon_{ijk}\,
	J_i\,\rho_R^j \wedge \rho_R^k,
	\nonumber \\
	&= -\sum_i dJ'_i \wedge \rho_L^i +
	\frac{1}{2}\sum_{ijk} \epsilon_{ijk} \,
	J'_i\,\rho_L^j \wedge\rho_L^k,
	\label{omegaQdot}
	\end{eqnarray}
where we have used (\ref{drhorels}).  The matrix of the components of
$\omega$ in the bases $(\rho_R^i,dJ_i)$ along the rows and
$(\rho_R^j,dJ_j)$ along the columns is 
	\begin{equation}
	\omega_{ab} =
	\left(\begin{array}{cc}
	\epsilon_{ijk}\,J_k & -\delta_{ij} \\
	\delta_{ij} & 0 \end{array}\right),
	\label{omegaab}
	\end{equation}
where $a,b=1,\ldots,6$.  Inverting this we obtain the matrix of the
Poisson tensor in the bases $(X^R_i,\partial/\partial J_i)$ along the
rows and $(X^R_j,\partial/\partial J_j)$ along the columns,
	\begin{equation}
	(\omega^{-1})^{ab} =
	\left(\begin{array}{cc}
	0 & \delta_{ij} \\
	-\delta_{ij} & \epsilon_{ijk}\,J_k 
	\end{array}\right)
	\label{Poissonab}
	\end{equation}
This in turn implies that the Poisson bracket of two functions on
$\Qdot\cong\TstarSU2dot$, when obtained by reduction from $\Phi_{2j}$,
is
	\begin{equation}
	\{F,G\} = \sum_i \left[(X^R_i \,F)\frac{\partial G}{\partial J_i}
	-\frac{\partial F}{\partial J_i}(X^R_i \,G)\right]
	+ \Jvec\cdot\left(
	\frac{\partial F}{\partial \Jvec} \times
	\frac{\partial G}{\partial \Jvec}\right).
	\label{PBQdot}
	\end{equation}
Carrying out the same procedure on the primed version of the
symplectic form, we obtain
	\begin{equation}
	\omega=-\sum_i dJ'_i \wedge \rho_L^i
	+\frac{1}{2} \sum_{ijk}\epsilon_{ijk}\,
	J'_i\,\rho_L^j\wedge\rho_L^k,
	\label{omegaQdot1}
	\end{equation}
and
	\begin{equation}
	\{F,G\} = \sum_i\left[
	-(X^L_i\,F)\frac{\partial G}{\partial J'_i}
	+\frac{\partial F}{\partial J'_i}(X^L_i \,G)\right]
	+ \Jvec'\cdot\left(
	\frac{\partial F}{\partial \Jvec'} \times
	\frac{\partial G}{\partial \Jvec'}\right).
	\label{PBQdot1}
	\end{equation}

To check these calculations (especially the signs) we can use the
Poisson brackets (\ref{PBQdot}) or (\ref{PBQdot1}) to study various
group actions on $\Qdot\cong\TstarSU2dot$.  For example, the
Hamiltonian $H=\nvec\cdot\Jvec$, where $\nvec$ is a unit vector,
produces the flow
	\begin{eqnarray}
	\frac{d\Jvec}{d\alpha} &= \nvec\times\Jvec, 
	\label{JnJeqn} \\
	\frac{dg}{d\alpha} &= \sum_i n_i (X_R^i\,g)
	=-\frac{i}{2} (\nvec\cdot\bsigma)g,
	\label{gnJeqn}
	\end{eqnarray}
where we use (\ref{PBQdot}) and (\ref{XLXRdef}) and where $\alpha$ is
the angle conjugate to $H$.  These have the solutions,
	\begin{equation}
	\Jvec \mapsto R(\nvec,\alpha)\Jvec,
	\qquad
	g \mapsto u(\nvec,\alpha)g,
	\label{nJflow}
	\end{equation}
where $u(\nvec,\alpha)$ is defined by (\ref{uaxisangle}) and where $R$
and $u$ are related by (\ref{Rijdef}).  Similarly, the flow generated
by $H=\nvec\cdot\Jvec'$ is 
	\begin{eqnarray}
	\frac{d\Jvec'}{d\alpha} &= \nvec\times\Jvec', 
	\label{JnJprimeeqn} \\
	\frac{dg}{d\alpha} &= \sum_i n_i (X^L_i\,g)
	=\frac{i}{2} g(\nvec\cdot\bsigma),
	\label{gnJprimeeqn}
	\end{eqnarray}
with solutions,
	\begin{equation}
	\Jvec' \mapsto R(\nvec,\alpha)\Jvec',
	\qquad
	g \mapsto gu(\nvec,\alpha)^{-1}.
	\label{nJprimeflow}
	\end{equation}
These are the expected actions of $SU(2)$ (left and right) on
$\TstarSU2dot$, given the actions (\ref{Jaction}) and
(\ref{Jprimeaction}) on $\Phi_{2j}$.  That is, one set is mapped into
the other through the relation $z=g\Theta z'$.

Another Poisson bracket of importance is 
          \begin{equation}
          \{J_i,J'_j\}=0,
          \label{JJprimePB}
          \end{equation}
one that is easiest to derive back on $\Phi_{2j}$, where $\Jvec$ is a
function of $z$ and $z^\dagger$, and $\Jvec'$ is a function of $z'$ and
$z^{\prime\dagger}$.  On $\TstarSU2dot$ the vanishing of this Poisson
bracket is a reflection of the fact that left and right translations
commute.  This Poisson bracket implies that $\Jvec'$ is constant along
the $\Jvec$-flow (\ref{nJflow}), and $\Jvec$ is constant along the
$\Jvec'$-flow (\ref{nJprimeflow}).

Another Hamiltonian function is of interest, namely,
$I=I'=|\Jvec|=|\Jvec'|$.  If $\alpha$ is the conjugate angle, then
Hamilton's equations are
       \begin{eqnarray}
       \frac{dg}{d\alpha} &= -\frac{i}{2J} (\Jvec\cdot\bsigma)g
       = +\frac{i}{2J}g(\Jvec'\cdot\bsigma)
       \label{Igeqn} \\
       \frac{d\Jvec}{d\alpha} &= \frac{d\Jvec'}{d\alpha} =0,
       \label{IJveceqn}
       \end{eqnarray}
where the two forms of the $g$-equation come from the two versions of
the Poisson bracket (\ref{PBQdot}) or (\ref{PBQdot1}), or through the
use of (\ref{SU2adjoint}) and (\ref{adjointeqn1}); and the $\Jvec$-
and $\Jvec'$- equations follow from (\ref{JJprimePB}).  These
equations have the solution,
       \begin{equation}
       g \mapsto u(\jvec,\alpha)g = gu(\jvec',\alpha)^{-1},
       \label{Iflow}
       \end{equation}
where $\jvec=\Jvec/J$ and $\jvec'=\Jvec'/J$ and where $\Jvec$ and
$\Jvec'$ are constant.  The orbit in $\TstarSU2dot$ is a circle upon
which $\Jvec$ and $\Jvec'$ are constant; the group is $U(1)$.

Finally, let us consider the Hamiltonian function
$\nvec\cdot(\Jvec+\Jvec')$.  Since $\Jvec$ and $\Jvec'$ commute, the
action is the product of the left and right actions shown in
(\ref{nJflow}) and (\ref{nJprimeflow}), that is,
	\begin{equation}
	\fl g\mapsto u(\nvec,\alpha)gu(\nvec,\alpha)^{-1},
	\qquad
	\Jvec \mapsto R(\nvec,\alpha)\Jvec,
	\qquad
	\Jvec'\mapsto R(\nvec,\alpha)\Jvec'.
	\label{JplusJprimeflow}
	\end{equation}

\subsection{Manifolds on the Quotient Space}
\label{manifoldsTstarSU2}

The $A$- and $B$-manifolds, defined as submanifolds of $\Phi_{12j}$ by
(\ref{Aset}) and (\ref{Bset}), are subsets of the level set $L$ of the
momentum map of the first reduction, that is, the conditions $I_r=I'_r$,
$r=1,\ldots,6$ hold on both the $A$- and $B$-manifolds.  Moreover, the
$A$- and $B$-manifolds are both invariant under the $U(1)^6$ group
action of that reduction, that is, they are composed of a union of
orbits of this group.  This means that they survive the projection
onto the quotient space $[\TstarSU2dot]^6$, so long as none of the
edge lengths is allowed to be zero (the condition $J_r>0$ was
understood in our original formulation of these manifolds).  The same
is true of the larger $\Atilde$- and $\Btilde$-manifolds, as long as
we exclude zero edge lengths.  On a single copy of $\TstarSU2dot$, the
condition $I=I'=|\Jvec|=|\Jvec'|$ is automatic, so on
$[\TstarSU2dot]^6$ the $A$-manifold is defined as the level set
$I_r=J_r$, $r=1,\ldots,6$, plus the four triangle conditions seen in
(\ref{Aset}), a total of $6+4\times3=18$ independent conditions that
specify an 18-dimensional Lagrangian manifold in $[\TstarSU2dot]^6$.

\begin{figure}[htb]
\begin{center}
\scalebox{0.43}{\includegraphics{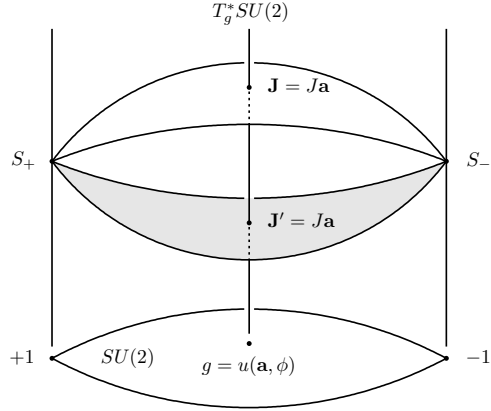}}
\end{center}
\caption[rps]{\label{LambdaJ} A schematic illustration of the
3-dimensional Lagrangian submanifold $\Lambda_J \in T^*SU(2)$ for
$J>0$.  The set $\Lambda_J$ consists of two 3-dimensional sections
over $SU(2) \setminus\{\pm1\}$, plus two 2-spheres $S_+$ and $S_-$
over $\pm1$.  Each section intersects the fiber $T^*_gSU(2)$ over
$g=u(\avec,\phi)\ne\pm1$ in one point; on the upper (unshaded) sheet,
the point is $\Jvec=J\avec=-\Jvec'$, while on the lower (shaded)
sheet, the point is $\Jvec=-J\avec=-\Jvec'$.}
\end{figure}

As for the $B$-manifold in $[\TstarSU2dot]^6$, it is the level set
$|\Jvec_r|=J_r$, $r=1,\ldots,6$ plus the six diangle conditions,
$\Jvec_r+\Jvec'_r=0$, $r=1,\ldots,6$, nominally $6+3\times6=24$
conditions.  However, these are not all independent.  We examine the
question of independence in a single copy of $\TstarSU2dot$, in which
we consider the submanifold defined by $\Jvec+\Jvec'=0$ and
$|\Jvec|=J$ where $J>0$ is given.  We call this submanifold $\Lambda_J
\subset \TstarSU2dot$.  Let $g\ne\pm1$ (1 means the identity in
$SU(2)$) be a group element, and consider the intersection of the
fiber $T_g^*SU(2)$ over $g$ with $\Lambda_J$.  Points of this fiber
are covectors at $g$, whose components with respect to the right- and
left-invariant bases are $\Jvec$ and $-\Jvec'$.  On the intersection
these components satisfy $\Jvec=-\Jvec'$ as well as
(\ref{adjointeqn1}).  Since we are assuming $\Jvec\ne0$,
(\ref{adjointeqn1}) can only be satisfied if $\Jvec=-\Jvec'$ is
parallel or antiparallel to the axis of the group element $g$.  That
is, if we write $g$ in axis-angle form, $g=u(\avec,\phi)$, where
$\avec$ is a unit vector, then we must have
	\begin{equation}
	\Jvec=\pm J \avec, \qquad \Jvec'=\mp J\avec.
	\label{JJprimeavec}
	\end{equation}
We exclude $g=\pm1$ because only for $g\ne\pm1$ is
the axis a unique unit vector in $\Reals^3$.  Therefore over
$SU(2)\setminus \{\pm1\}$ the manifold $\Lambda_J$ consists of two
sections of the bundle $T^*SU(2)$, so it is 3-dimensional.  It is also
Lagrangian, being the level set of functions that commute on the level
set.  As for the points $g=\pm1$, here $R(g)=I$ (the identity
$3\times3$ rotation) and the intersection of $T_g^*SU(2)$ with
$\Lambda_J$ is the 2-sphere $|\Jvec|=|\Jvec'|=J\ne0$.  This means that
the Lagrangian manifold $\Lambda_J$ is vertical in two dimensions over
$g=\pm1$ (these points are second order caustics).  The two branches of
$\Lambda_J$ over $SU(2)\setminus\{\pm1\}$ merge together at $g=\pm1$,
and $\Lambda_J$ is connected.  The manifold $\Lambda_J$ for $J>0$ is
illustrated in Fig.~\ref{LambdaJ}.  Finally, although the set $J=0$ (that
is, $\Jvec=-\Jvec'=0$) is not a part of $\TstarSU2dot$ it does belong
to $T^*SU(2)$; it is just the zero section of the bundle, which is
diffeomorphic to $SU(2)$ itself, and is therefore 3-dimensional.  The
manifolds $\Lambda_J$ for $J\ge0$ form a 1-parameter family of
3-dimensional Lagrangian manifolds in $T^*SU(2)$.

Finally, the $B$-manifold is the six-fold product $\Lambda_{J_1}
\times \ldots \times \Lambda_{J_6}$; it is an 18-dimensional
Lagrangian submanifold of the quotient space $[\TstarSU2dot]^6$. 

As for the enlarged manifolds $\Atilde$ and $\Btilde$, if we exclude
zero edge lengths then these project onto $24$-dimensional
submanifolds of $[\TstarSU2dot]^6$.  Manifold $\Atilde$ is specified
simply by the four triangle conditions, that is $4\times3=12$ scalar
conditions.  As for manifold $\Btilde$, on a single copy of
$\TstarSU2dot$ let us define
	\begin{equation}
	\Lambda =\bigcup_{J>0} \Lambda_J,
	\label{Lambdadef}
	\end{equation}
which we can extend to $J\ge0$ on $T^*SU(2)$ when relevant.  Then
$\Lambda$ is a 4-dimensional submanifold of $\TstarSU2dot$ or
$T^*SU(2)$, and the manifold $\Btilde$ is the 24-dimensional subset of
$[\TstarSU2dot]^6$ given by $\Lambda^6$.  It is also the level set of
the diangle conditions $\Jvec_r+\Jvec'_r=0$, $r=1,\ldots,6$; these are
$3\times6=18$ conditions, the same ones that define $\Btilde$ as a
subset of $\Phi_{12j}$, but only 12 are independent on $[T^*SU(2)]^6$,
as indicated by the fact that $\Btilde$ is 24-dimensional.

Now we return to Fig.~\ref{ABintsect} and consider the projection of
the manifolds illustrated there onto the quotient space
$[\TstarSU2dot]^6$.  The projections of the $A$- and $B$-manifolds
have just been described; each loses six dimensions upon projection,
becoming 18-dimensional, because each is a union of 6-dimensional
orbits of the symmetry group $U(1)^6$.  The same applies to their
intersections $T_1$ and $T_2$, which drop from 15 dimensions to 9.
The (projected) 9-dimensional versions of $T_1$ and $T_2$ on
$[\TstarSU2dot]^6$ are the orbits of a 9-dimensional common group,
which is worth describing.

The momentum map for the common group consists of the six functions
$I_r=|\Jvec_r|=|\Jvec'_r|$ on $[\TstarSU2dot]^6$, as well as the three
functions in $\Jvec_{\rm tot}$, defined by (\ref{Jtotdef}).  The group
itself is $U(1)^6 \times SU(2)$.  As for the $I_r$, the flow of each
one acts on only one copy of $\TstarSU2dot$, and that action was
presented in (\ref{Iflow}) and below.  Thus the set of the six $I_r$
generates an action of $U(1)^6$ with orbits that are 6-tori, upon
which the vectors $\Jvec_r$ and $\Jvec'_r$ are constant.  Since the
diangle and triangle conditions depend only on these vectors, if they
are satisfied at one point on the 6-torus, they are satisfied at all
points on it; thus, the projected $T_1$ and $T_2$ manifolds in
$\TstarSU2dot$ are unions of such orbits.  As for $\Jvec_{\rm tot}$,
this is a sum over $r$ of vectors of the form $\Jvec+\Jvec'$ for each
$r$; the flow generated by each one of these is given by
(\ref{JplusJprimeflow}).  The effect of the $SU(2)$ group action is
thus to conjugate each $g_r$ by the same element of $SU(2)$, while
rotating all of the vectors $\Jvec_r$ and $\Jvec'_r$ by the
corresponding rotation in $SO(3)$.  The latter leaves the diangle and
triangle conditions invariant, thus maintaining the conditions for a
tetrahedron.  Thus, the projected manifolds $T_1$ and $T_2$ are
invariant under this action.  Assembling these facts, one can show
that the manifolds $T_1$ and $T_2$ in $\TstarSU2dot$ consist of
precisely one 9-dimensional orbit of the group $U(1)^6 \times SU(2)$
(for the assumed values of the $J_r$).

Finally, the contour $P\to Q\to P'\to P$ illustrated in
Fig.~\ref{ABintsect} projects onto a similar contour in
$[\TstarSU2dot]^6$.  Also, since the symplectic 1-form on
$\Phi_{12j}=\Phi_{2j}^6$ is the pull-back of that on
$[\TstarSU2dot]^6$, the integrals of the symplectic 1-forms along the
respective contours are the same.  (In general in symplectic reduction,
it is only the symplectic 2-forms that are guaranteed to be related by
the pull-back, but in this example it works for the 1-forms.)  This
means that the integral representation of the Ponzano-Regge phase
presented in Sec.~\ref{pathPRphase}) projects onto
$[\TstarSU2dot]^6$. 

Similarly, the manifolds illustrated in Fig.~\ref{iron} also project
onto $[\TstarSU2dot]^6$, losing six dimensions in the process.  The
proof of the Schl\"afli identity given in Sec.~\ref{proof} and based on
these manifolds also carries over to the quotient space with just a
change of context.  One of the manifolds used in this proof is
$\Ttilde$, which is the union of all manifolds $T_1$ and $T_2$ for all
values of the $J_r$ (we may exclude $J_r=0$).  This is the manifold of
tetrahedra, and being a 6-parameter family of 9-dimensional manifolds,
it is 15-dimensional.

\subsection{Semiclassical Interpretation of $\Lambda_J$}
\label{semiclassinterp}

We make some final remarks of a semiclassical nature concerning the
Lagrangian manifolds $\Lambda_J \subset T^*SU(2)$.  Whenever a
Lagrangian manifold occurs in a classical context it is useful to ask
what its semiclassical significance is.  In the case of the
$\Lambda_J$ it is easiest to answer this question with respect to the
version of $T^*SU(2)$ obtained from $\Phi_{2j}^*$ rather than from
$\Phi_{2j}$.  In the following we will proceed somewhat intuitively,
stating several plausible facts without proof and switching back and
forth between the classical geometry and the corresponding objects and
operations in the linear algebra of the corresponding quantum systems.

For a given $J$, $\Lambda_J\subset T^*SU(2)$ is the 3-dimensional
level $I=|\Jvec|=J$, $\Jvec=\Jvec'$.  Let us denote its preimage under
the projection map $:\Ldot\to\TstarSU2dot$ by $M_J$, which is the
level set $I=J$, $\Jvec=\Jvec'$ inside $\Phi_{2j}^*$.  Then
$M_J\subset\Ldot\subset\Phi_{2j}^*$, and $M_J$ is a 4-dimensional
Lagrangian manifold in $\Phi_{2j}^*$.  It will be easiest to explain
the semiclassical significance of $M_J$ first, and then to turn to
$\Lambda_J$.

We denote the Hilbert space of two harmonic oscillators by $\Hspace$
as above.  As explained by Schwinger (1952) and Bargmann (1962),
$\Hspace$ carries a unitary representation of $SU(2)$ which we denote
by $g\mapsto U(g)$, and under this action $\Hspace$ decomposes into
one carrier space of irrep $j$ for each $j$,
	\begin{equation}
	\Hspace=\bigoplus_j \Hspace_j.
	\label{Hjdef}
	\end{equation}
A convenient basis on $\Hspace$ is $\{\ket{jm},
\forall j,m\}$, where ``$\forall j,m$'' means all $j$ and $m$ allowed
by the usual restrictions on angular momentum quantum numbers.  A
basis in $\Hspace_j$ is the restricted set $\{\ket{jm}, \forall m\}$.

The Hilbert space corresponding to $\Phi_{2j}^*$ is
$\Hspace\otimes\Hspace^*$, that is, it consists of linear combinations
of vectors of the form $\ketbra{\phi}{\psi}$, in Dirac notation.
These vectors are otherwise linear operators $:\Hspace\to\Hspace$.  A
convenient basis on $\Hspace\otimes\Hspace^*$ is the set
$\{\ketbra{jm}{j'm'}, \forall j,j',m,m'\}$.  Now the subset of
$\Phi_{2j}^*$ indicated by $I=I'$, that is, the set
$L\subset\Phi_{2j}^*$, corresponds to the subspace of
$\Hspace\otimes\Hspace^*$ on which $j=j'$.  Let us call this space
$\Lspace \subset \Hspace\otimes\Hspace^*$.  Space $\Lspace$ is spanned
by $\{\ketbra{jm}{jm'}, \forall j,m,m'\}$, and it contains the
operators $:\Hspace\to\Hspace$ that leave each irreducible subspace
$\Hspace_j$ invariant.  That is, the operators in $\Lspace$ are are
block diagonal in the $\ket{jm}$ basis.  In summary, $\Lspace$ is the
space of operators corresponding semiclassically to the manifold $L$.

Restricting the level set further by requiring that $I=I'=J$ for a
given $J$, we obtain a classical submanifold of $\Phi_{2j}^*$ that
corresponds to operators that are linear combinations of the set
$\{\ketbra{jm}{jm'}, \forall m,m'\}$, that is, operators in $\Lspace$
with a fixed value of $j$.  These operators map $\Hspace_j$ into
itself for the given $j$, while annihilating all vectors in
$\Hspace_{j'}$ for $j'\ne j$.  Let us call this space $\Lspace_j
\subset \Lspace$; it has dimensionality $(2j+1)^2$.

Restricting the level set $I=I'=J$ further by adding $\Jvec=\Jvec'$,
we obtain the classical manifold $M_J$.  On the Hilbert space
$\Hspace\otimes\Hspace^*$ the operators $\Jvec-\Jvec'$ generate an
action of $SU(2)$ given by $X\mapsto U(g)XU(g)^\dagger$, where
$X:\Hspace\to\Hspace$ is any operator.  Also, the condition
$\Jvec-\Jvec'=0$ specifies a subspace of operators for which
$U(g)XU(g)^\dagger=X$, that is, operators that are invariant under
rotations.  But the only operator in $\Lspace_j$ that is invariant
under rotations is the projection operator onto subspace $\Hspace_j$,
	\begin{equation}
	\Pi_j = \sum_m \ketbra{jm}{jm},
	\label{Pijdef}
	\end{equation}
according to Schur's lemma.  This is the operator or vector in
$\Lspace_j \subset \Lspace \subset \Hspace\otimes\Hspace^*$ that
corresponds to the classical Lagrangian manifold $M_J$.

Now we turn to the projection of $L$ onto $T^*SU(2)$ and of $M_J
\subset L$ onto $\Lambda_J \subset T^*SU(2)$, and explain the
corresponding linear algebra.  The symplectic manifold $T^*SU(2)$ is
the cotangent bundle of $SU(2)$, so it should be the phase space
corresponding to $L^2(SU(2))$, the Hilbert space of wave functions on
$SU(2)$.  An obvious basis on this Hilbert space is the set of
components of the rotation matrices, $\{D^j_{mm'}(g), \forall
j,m,m'\}$, which are orthonormal with respect to the Haar measure on
the group.  Actually, it is better to use the complex conjugates of
these functions on $SU(2)$, $\overline{D^j_{mm'}(g)}$, which, when
regarded as wave functions of a rigid body, have the correct
transformation properties under rotations (Littlejohn and Reinsch
1997).  In view of the quantum numbers, these wave functions are
obviously the images of the vectors $\ketbra{jm}{jm'}$ under the
quantum analog of the classical projection map, and it would appear
that we have a map $:\Lspace\to L^2(SU(2))$.  

Explicitly, this map is
	\begin{equation}
	X\mapsto \psi_X(g) = \tr[U(g)^\dagger X],
	\label{psiXdef}
	\end{equation}
where $X\in\Lspace$.  In particular, the basis vector
$\ketbra{jm}{jm'}$ is mapped into
	\begin{equation}
	\tr[U(g)^\dagger \ketbra{jm}{jm'}] =
	\matrixelement{jm'}{U(g)^\dagger}{jm} =
	\overline{D^j_{mm'}(g)},
	\label{jmmtoD}
	\end{equation}
as expected.  As for the projection operator $\Pi_j$, its wave
function on $SU(2)$ is the character,
	\begin{equation}
	\Pi_j \mapsto \tr[U(g)^\dagger \Pi_j] = \sum_m
	\overline{D^j_{mm}(g)} = \chi^j(g),
	\label{Pijchij}
	\end{equation}
where $\chi^j(g)$ actually depends only on the conjugacy class, and
where we have used the fact that for $SU(2)$ the characters are real.
Thus we see that the wave function on $SU(2)$ corresponding to the Lagrangian
manifold $\Lambda_J$ is the character $\chi^j(g)$.

The trace in (\ref{psiXdef}) is the scalar product in
$\Hspace\otimes\Hspace^*$, so semiclassically its stationary phase
points should be the intersections of the Lagrangian manifolds in
$\Phi_{2j}^*$ corresponding to operators $X$ and $U(g)$.  As for
$U(g)$, its Lagrangian manifold is $z=gz'$, precisely the coordinate
transformation we used in Sec.~\ref{reductionPhi2jstar} on passing
from $\Ldot$ to $\TstarSU2dot$.  One way to see this is to note that
the graph of a symplectic map $:\Phi\to\Phi$ in $\Phi\times\Phi^*
=\Phi_{2j}^*$ is a Lagrangian manifold in the latter space.  In this
case, $z=gz'$, which is the obvious symplectic map corresponding to
$U(g)$, specifies a Lagrangian manifold in $\Phi_{2j}^*$ that is the
obvious candidate for the manifold supporting the operator $U(g)$.
The fact that it is Lagrangian can be verified directly by computing
the Poisson brackets among the components of $z-gz'$ and
$z^\dagger-z^{\prime\dagger}g^\dagger$ in the symplectic structure
(\ref{omega2jstar}).  These Poisson brackets all vanish.  The
Lagrangian manifold $z-gz'=0$ in $\Phi_{2j}^*$ is actually a plane,
which moreover is a subset of $L$.  

Another point of view is the following.  The operator $U(g)$ belongs
to the space $\Hspace\otimes\Hspace^*$.  Let us call operators in this
space ``ordinary operators,'' while linear operators mapping this
space into itself we will call ``superoperators'' (terminology without
relation to supersymmetry).  Then an ordinary operator can be
associated with a superoperator by either left or right
multiplication.  For example, the ordinary operator $a_\mu$, the usual
annihilation operator, becomes a superoperator by either left or right
multiplication, $X\mapsto a_\mu X$ or $X \mapsto Xa_\mu$, where $X$ is
an ordinary operator.  Then on the phase space $\Phi_{2j}^*$ the
superoperator, left multiplying by $a_\mu$, has Weyl symbol $z_\mu$,
while the superoperator, right multiplying by $a_\mu$, has Weyl symbol
$z'_\mu$, in the $(z,z')$ coordinates we have been using on
$\Phi_{2j}^*$.  These operators satisfy the relation,
	\begin{equation}
	a_\mu U(g) = U(g) \sum_\nu g_{\mu\nu}\,a_\nu,
	\label{aUcommrel}
	\end{equation}
which can be expressed by saying that $U(g)$ is an eigenoperator of
the commuting superoperators whose symbols are the components of
$z-gz'$ with eigenvalues 0.  Thus $U(g)$ is supported semiclassically
by the Lagrangian manifold $z=gz'$.  

The Lagrangian manifold $z=gz'$ in $\Phi_{2j}^*$ is a subset of $L$
and moreover consists of orbits of $I-I'$.  Thus it survives the
projection onto the quotient space, where in fact it becomes simply
$g={\rm const}$ on $T^*SU(2)$.  That is, it is the fiber over the
point $g$ in the cotangent bundle.  This fiber is naturally
Lagrangian.  So the intersections of Lagrangian manifolds implied by
(\ref{psiXdef}) projects onto an intersection of Lagrangian manifolds
in $T^*SU(2)$, one being $\Lambda_J$ and the other the fiber over $g$
in $T^*SU(2)$.  This is the expected geometry for a wave function on
$SU(2)$.

Finally we note that the characters in $SU(2)$ are given explicitly by
	\begin{equation}
	\chi^j(\phi) = \frac{\sin(j+1/2)\phi}{\sin(1/2)\phi},
	\label{chij}
	\end{equation}
where we write $g$ in axis-angle form, $g=u(\avec,\phi)$.  By tracking
the Lagrangian manifolds and the densities on them through the
projection process we have just described it can be shown that the
denominator $\sin(1/2)\phi$ can be interpreted as an amplitude
determinant in the semiclassical expression for the wave function.
The amplitude diverges at $\phi=0$ and $\phi=2\pi$, that is, at the
group elements $g=\pm1$, where as we have noted the Lagrangian
manifold has caustics when projected onto the base space $SU(2)$.  The
numerator $\sin(j+1/2)\phi$ is the sum of two branches of a WKB wave
function, corresponding to the two branches of $\Lambda_J$ seen in
Fig.~\ref{LambdaJ} or in Fig.~\ref{Jcircle} below.  

In the case of the character formula the semiclassical treatment is
exact.  This can be seen as an example of the Duistermaat-Heckman
theorem, which also connects this discussion with mathematical work on
the asymptotics of group structures.  See Duistermaat and Heckman
(1982), Thompson and Blau (1997), Stone (1989) and Ben Geloun and
Gurau (2011).  

\section{The Second Reduction}
\label{secondreduction}

We now carry out a second reduction, this one taking us from
$[\TstarSU2dot]^6$ or $[T^*SU(2)]^6$ to a new quotient space we
denote by $\Sigmadot$ or $\Sigma$.  It will turn out that $\Sigma$ is
essentially the symplectic manifold found in an {\it ad hoc} manner in
Sec.~\ref{LMoftetrahedra} and illustrated in Fig.~\ref{schlafliL}.
The symmetry group for this reduction is $SU(2)^6$, and its momentum
map and the level set to be used are given by $\Jvec_r+\Jvec'_r=0$,
$r=1,\ldots,6$.

Before we develop this further, however, there is one obvious problem,
if we intend to project the contour for the Ponzano-Regge phase,
illustrated in Fig.~\ref{ABintsect}, onto $\Sigmadot$.  That is,
although the $B$-manifold is a subset of the level set
$\Jvec_r+\Jvec'_r=0$, $r=1,\ldots,6$, as required if it is to
survive the projection onto $\Sigmadot$, the same does not apply to
the $A$-manifold.  Therefore we do not seem to have a contour for the
Ponzano-Regge phase on $\Sigmadot$.  We address this problem before
studying the second reduction further.

\subsection{A New Contour for the Ponzano-Regge Phase}
\label{newcontour}

Consider the contour $P\to Q\to P'\to P$ illustrated in
Fig.~\ref{ABintsect}.  It was explained in Sec.~\ref{PRphase} that the
integral of the symplectic 1-form $\theta$ around this contour is
$2S$, where $S$ is the Ponzano-Regge phase given by (\ref{Sdef}).  In
this section we interpret this contour as belonging to either
$\Phi_{12j}$ or $\TstarSU2dot\subset T^*SU(2)$ and $\theta$ as given by
either (\ref{theta12j}) or (\ref{thetaQdot3}).  However, it was also
pointed out in Sec.~\ref{PRphase} that the integrals of $\theta$ along
the legs $P\to Q$ and $Q\to P'$ vanish, so in fact the closed contour
$P\to Q\to P'\to P$ can be restricted to just the open contour $P' \to
P$ without changing the result.  The latter contour is an open contour
contained in the intersection set $T_1$.

\begin{figure}[htb]
\begin{center}
\scalebox{0.43}{\includegraphics{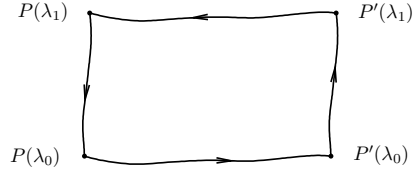}}
\end{center}
\caption[rps]{\label{leg3}A reduced contour $C$ for the Ponzano-Regge
phase.  The contour lies entirely inside the set $\Ttilde$ of 
tetrahedra.}
\end{figure}  

Next, in Sec.~\ref{proof} a family of such contours, parameterized by
$\lambda$, was considered, as illustrated in Fig.~\ref{iron}.  These
sweep out a 2-dimensional surface illustrated in the figure.  Applying
Stokes' theorem to the integral of $\omega=d\theta$ over this surface
(the ``walls''), we find that this integral is $2[S(\lambda_1)
-S(\lambda_0)]$.  On the other hand, it was pointed out in
Sec.~\ref{proof} that the integrals of $\omega$ over the first and
second legs of the surface vanish, as indicated by (\ref{wallint}).
Now applying Stokes' theorem to the integral over the third leg only, we
obtain
	\begin{equation}
	2[S(\lambda_1)-S(\lambda_0)] = \int_C \theta,
	\label{PRreducedcontour}
	\end{equation}
where $C$ is the closed contour $P(\lambda_0) \to P'(\lambda_0) \to
P'(\lambda_1) \to P(\lambda_1) \to P(\lambda_0)$ which may be seen in
Fig.~\ref{iron} and which is illustrated separately in Fig.~\ref{leg3}.
In fact, the integral along the top and bottom edges of $C$ give the
result shown in (\ref{PRreducedcontour}), so the integrals of $\theta$
along the vertical segments in Fig.~\ref{leg3} must cancel each other.

For our purposes the main point is that this new contour $C$ lies
entirely in the space $\Ttilde$ of tetrahedra. This is a subset of
$\Btilde$, which survives the projection onto $\Sigmadot$, so the
contour $C$ illustrated in Fig.~\ref{leg3} does also, and gives us an
integral representation of the Ponzano-Regge phase (actually the
difference seen in (\ref{PRreducedcontour})) on the quotient space
$\Sigmadot$.  

Before leaving this contour we will specialize the initial conditions
for future convenience.  The 2-dimensional surface illustrated in
Fig.~\ref{iron} is swept out by curves $P(\lambda) \to Q(\lambda) \to
P'(\lambda) \to P(\lambda)$.  All we require of the initial point
$P(\lambda)$ is that it lie in the space $\Ttilde$ of tetrahedra, that
is, that the triangle and diangle conditions be satisfied.  Supposing
this to be true, there still remains the question of the initial phases
of the spinors $(z_r,z'_r)$.  Let us now assume that at $P(\lambda)$,
$z'_r = \zeta_r$, where $\zeta_r$ is a spinor such that
$(1/2)\zeta_r^\dagger \bsigma \zeta_r=\Jvec'_r$.  Let us also assume
that at $P(\lambda)$, $z_r=\Theta\zeta_r$.  This implies
$(1/2)z_r\bsigma z_r = \Jvec_r=-\Jvec'_r$, as required of the initial
conditions.  

Then the procedure of Sec.~\ref{PRphase} shows that at point
$P'(\lambda)$, we have $z'_r=\zeta_r$ (the same as at $P(\lambda)$),
and $z_r = e^{i\psi_r} \Theta\zeta_r$, where $\psi_r=\psi_r(\lambda)$
is the dihedral angle of edge $r$ of the initial tetrahedron.  Thus,
one can say that the contour $C$ illustrated in Fig.~\ref{leg3} has
the dihedral angles built into it.

These choices are convenient, because they imply that at
$P(\lambda)$ (and therefore all along the segment $P(\lambda_1) \to
P(\lambda_0)$ in Fig.~\ref{leg3}) we have $g_r=1$, $r=1,\ldots,6$.  As
for point $P'(\lambda)$, we have 
	\begin{equation}
	g_r = u(\jvec_r,-2\psi_r)= u(-\jvec_r,2\psi_r),
	\label{gratPprime}
	\end{equation}
because this $SU(2)$ element, when applied to $\Theta\zeta_r$, brings
out the phase $e^{i\psi_r}$.  Here $\jvec_r=\Jvec_r/J_r$.    

\subsection{Back to the Second Reduction}

We now carry out the second reduction, in which the group is
$SU(2)^6$, generated by $\Jvec_r+\Jvec'_r$, $r=1,\ldots,6$.  We work
on the level set $\Jvec_r+\Jvec'_r=0$, which is otherwise $\Btilde$.
For simplicity we work with the space $[T^*SU(2)]^6$ rather than
$[\TstarSU2dot]^6$.  As before the group action is a product of an
action on each factor, in this case $SU(2)$ acting on $T^*SU(2)$; the
latter action is given explicitly by (\ref{JplusJprimeflow}).  We drop
the $r$ index when considering a single factor.

The zero level set in the 6-dimensional space $T^*SU(2)$ is $\Lambda$,
defined by (\ref{Lambdadef}); it is 4-dimensional.  When considering
the action of $SU(2)$ on $\Lambda$ it is convenient to decompose
$\Lambda$ into the family of 3-dimensional submanifolds $\Lambda_J$,
as indicated by (\ref{Lambdadef}), since each $\Lambda_J$ is invariant
under the same action.  See Fig.~\ref{LambdaJ}.  To show this explicitly,
consider a point $g=u(\avec,\phi)\ne\pm1$ as in the figure, so that
the action (\ref{JplusJprimeflow}) implies
	\begin{equation}
	g=u(\avec,\phi) \mapsto
	u(\nvec,\alpha)u(\avec,\phi)u(\nvec,\alpha)^{-1}
	=u(\bvec,\phi),
	\label{innerauto}
	\end{equation}
where $\bvec=R(\nvec,\alpha)\avec$.  That is, the axis of the rotation
transforms the same as $\Jvec$ and $\Jvec'$, so the conditions
$\Jvec=\pm J\avec$, $\Jvec'=\mp J\avec$ are maintained by the action
(\ref{JplusJprimeflow}).  As for the points $g=\pm1$, the 2-spheres
over these points (sets $S_+$ and $S_-$ in the diagram) are invariant
under the action (\ref{JplusJprimeflow}).  It is also easy to see that
$\Lambda_J$ for $J=0$ is invariant under the action.  Since all sets
$\Lambda_J$ are invariant under the action, so is $\Lambda$, and the
isotropy subgroup is the entire group $SU(2)$.

As for the orbits of the action, consider first $J>0$ and $g\ne\pm1$.
Then the action (\ref{JplusJprimeflow}) causes the point $g$ to sweep
out a conjugacy class in $SU(2)$, which is a 2-sphere; and the point
on $\Lambda_J$ over $g$ on either (upper or lower) section sweeps
out the portion of the section over the conjugacy class.  But since
proper rotations cannot connect $\Jvec=J\avec$ with $\Jvec=-J\avec$,
the orbits of the action (\ref{JplusJprimeflow}) lie entirely on the
upper or lower sections, and the two branches are not connected by the
action.  Thus the orbits are copies of the conjugacy class over which
they lie, and are 2-spheres.  Similarly, over the points $g=\pm1$
the orbits are 2-spheres, so all orbits for $J>0$ are 2-spheres.  As
for the set $\Lambda_J$ for $J=0$, the orbits are the conjugacy
classes, which are 2-spheres except at $g=\pm1$ where they are single
points. 

\begin{figure}[htb]
\begin{center}
\scalebox{0.43}{\includegraphics{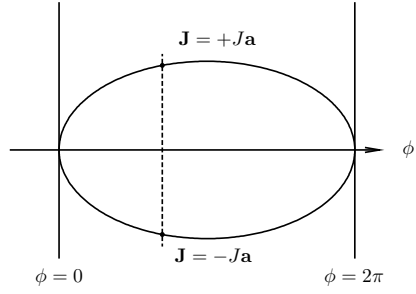}}
\end{center}
\caption[rps]{\label{Jcircle}Illustration of the quotient space
$\Lambda_J/SU(2)$ for the case $J>0$.  The space is a circle, which
can be regarded as a two-branched object over the space $0<\phi<2\pi$
of conjugacy classes, with the two branches joining at $\phi=0$ and
$\phi=2\pi$. The upper branch represents
the part of the cotangent bundle $T^*SU(2)$ over the conjugacy class
for which $\Jvec=+J\avec$ and $\Jvec'=-J\avec$, while on the lower
branch $\Jvec=-J\avec$ and $\Jvec'=+J\avec$.}
\end{figure}  

To parameterize the conjugacy classes of $SU(2)$ we write
$g=u(\avec,\phi)$ so that $\phi$ in the range $0\le\phi\le 2\pi$ is a
coordinate of the conjugacy classes, with $\phi=0$ meaning $g=1$ and
$\phi=2\pi$ meaning $g=-1$.  Then to construct the quotient space we
take first the case $J>0$, for which $\Lambda_J$ is 3-dimensional and
all orbits are 2-spheres, so that the quotient space is 1-dimensional.
It consists of two branches over $0<\phi<2\pi$, corresponding to
$\Jvec=\pm J\avec$, which join together in a single branch over
$\phi=0$ and $\phi=2\pi$, as illustrated in Fig.~\ref{Jcircle}.  The
quotient space $\Lambda_J/SU(2)$ for $J>0$ is a circle.  We introduce
a coordinate $\tau$ on this circle by defining $\tau=\phi$ on the
upper branch where $\Jvec=J\avec$, and $\tau=-\phi$ on the lower
branch where $\Jvec=-J\avec$.  Then we can take $\tau$ in the range
$-2\pi<\tau\le 2\pi$.  This takes care of the case $J>0$.

Forming the union of these circles for all $J>0$, we obtain a cylinder
with coordinates $(J,\tau)$, $0<J<\infty$.  This is the quotient
manifold under the second symplectic reduction applied to
$\TstarSU2dot$.  At $J=0$ the set $\Lambda_J$ is just $SU(2)$ itself,
and the quotient space is the one-dimensional interval
$0\le\phi\le2\pi$, the set of conjugacy classes.  Therefore we can
pinch off the cylinder at $J=0$ into this interval to form the entire
quotient space.  Alternatively, since the set $J=0$ was a single point
of the level set in the first reduction and the set $\Lambda_J$ for
$J=0$ is not a part of the image of the first reduction, it is
reasonable to take care of the case $J=0$ in an ad hoc manner by
pinching off the cylinder at $J=0$ into a point.  The resulting set is
topologically $\Reals^2$, with polar coordinates $(J,\tau)$ with
$-2\pi<\tau\le2\pi$.  Recall also that the case $J=0$ means a
tetrahedron for which the dihedral angles are not defined, so it plays
no role in the Schl\"afli identity.

Extending this construction to all values of $r=1,\ldots,6$, we can
define the quotient space $\Sigmadot$ as the sixth power of the
cylinder with coordinates $(J,\tau)$ just described; and similarly, we
define $\Sigma$ as the  sixth power of the plane $\Reals^2$ with polar
coordinates $(J,\tau)$.

\subsection{The Symplectic Form on $\Sigma$ or $\Sigmadot$}
\label{thetaSigmadot}

To compute the symplectic form on the $(J,\tau)$ cylinder we first
restrict the form (\ref{thetaQdot2}) to the level set, upon which
$\Jvec=\pm J\avec$.  This gives
	\begin{equation}
	\theta= i\Jvec\cdot\tr(\bsigma\,dg\, g^\dagger)
	= \pm iJ \tr[dg\,g^\dagger(\avec\cdot\bsigma)].
	\label{thetaJtau}
	\end{equation}
But $g=\cos\phi/2-i(\avec\cdot\bsigma)\sin\phi/2$, so
	\begin{equation}
	g^\dagger(\avec\cdot\bsigma) =
	(\avec\cdot\bsigma)\cos\phi/2 + i\sin\phi/2.
	\end{equation}
Also,
	\begin{equation}
	dg=\frac{1}{2}[-\sin\phi/2-i(\avec\cdot\bsigma)
	\cos\phi/2]d\phi - i(d\avec\cdot\sigma)
	\sin\phi/2.
	\label{dgeqn}
	\end{equation}
Now upon substituting (\ref{dgeqn}) into (\ref{thetaJtau}) the term
involving $d\avec$ vanishes, since $\avec\cdot d\avec=0$ ($\avec$ is a
unit vector) and since $\tr\bsigma=0$.  As for the term in $d\phi$, it
gives
	\begin{equation}
	\theta=\pm J\,d\phi,
	\end{equation}
after a bit of algebra, where the $\pm$ sign refers to the upper or
lower branches.  But this can be expressed in terms of the coordinate
$\tau$ on the circles, giving
	\begin{equation}
	\theta=J\,d\tau.
	\label{thetacylinder}
	\end{equation}
This is the symplectic form on the cylinder, or the plane $\Reals^2$,
if we extend it to $J=0$ in the manner suggested above.  It looks like
the symplectic form for the harmonic oscillator in action-angle
variables, except that the angle $\tau$ has a range of $4\pi$.   

Taking the sixth power of this, we obtain the symplectic 1-form on
$\Sigma$ or $\Sigmadot$, which is
	\begin{equation}
	\theta=\sum_r J_r\,d\tau_r.
	\label{thetaSigma}
	\end{equation}

\begin{figure}[htb]
\begin{center}
\scalebox{0.43}{\includegraphics{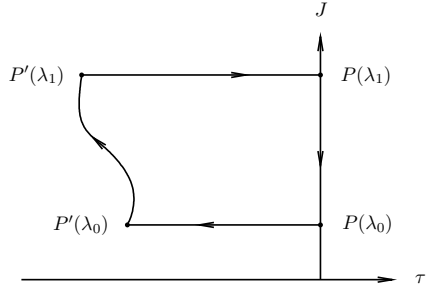}}
\end{center}
\caption[rps]{\label{Jtaucontour}When the contour illustrated in
Fig.~\ref{leg3} is projected onto the quotient space $\Sigmadot$, and
then projected again onto one of the $J$-$\tau$ spaces for some $r$,
it becomes the contour illustrated here.  On the upper segment
$J=J(\lambda_1)$ and on the lower $J=J(\lambda_0)$, while on the
vertical segment at the left, $\tau=-2\psi(\lambda)$ and
$J=J(\lambda)$.}
\end{figure}  

\subsection{The Contour and Lagrangian Manifold on $\Sigmadot$}
\label{contourSigmadot}

Now let us project the contour illustrated in Fig.~\ref{leg3} onto the
reduced space.  The contour contains a path $P(\lambda_1) \to
P(\lambda_0)$, along which $\Jvec_r$ and $\Jvec'_r$ are functions of
$\lambda$, defining a family of tetrahedra for which (by our
assumptions) $V>0$; therefore, along this path, $J_r$ and $\psi_r$ are
functions of $\lambda$, with $J_r>0$.  Also, according to the
conventions introduced in Sec.~\ref{newcontour}, we have $g_r=1$ along
this path, and hence $\tau_r=0$.  Next, along the path $P(\lambda_0)
\to P'(\lambda_0)$, $J_r$ is constant while the group element $g_r$
has axis $\avec=-\jvec_r$ and an angle (or conjugacy class) $\phi$
that goes from 0 to $2\psi_r$, as indicated by (\ref{gratPprime}).
This means that $\Jvec=-J\avec$, so we are on the lower branch of
Fig.~\ref{Jcircle}, and $\tau_r$ goes from 0 to $-2\psi_r(\lambda_0)$.
The path $P'(\lambda_1)\to P(\lambda_1)$ is similar; $J_r$ is constant
along this path, and $\tau_r$ goes from $-2\psi_r(\lambda_1)$ to 0.
Finally, along the path $P(\lambda_0) \to P(\lambda_1)$, $J_r$ and
$\tau_r$ are functions of $\lambda$, with $\tau_r =
-2\psi_r(\lambda)$.  This projected contour lies in the space
$\Sigmadot$ or $\Sigma$; when it is projected onto one of the
$(J,\tau)$ spaces for a fixed $r$, we obtain Fig.~\ref{Jtaucontour}.

Finally, the manifold on $\Sigmadot$ given by $\tau_r = -2\psi_r$,
where the $\psi_r$ are functions of the six $J_r$'s, is a Lagrangian
manifold, according to the logic of Sec.~\ref{LMoftetrahedra}.  In
fact, $\Sigmadot$ is essentially the symplectic manifold constructed
in Sec.~\ref{LMoftetrahedra}, apart from signs and factors of 2.  A
surprising aspect of this result is that the dihedral angles of the
tetrahedron emerge in the end as labels of conjugacy classes in
$SU(2)$.  

\section{Conclusions}
\label{conclusions}

To conclude we will mention some aspects of this calculation that have
not been mentioned so far.  First, it would be possible to carry out
the two symplectic reductions of this paper in a single step, that is,
by taking the momentum map and level set to be $\Jvec_r+\Jvec'_r=0$,
specifying the first and only reduction.  This condition implies
$I_r=I'_r$, so it includes the first reduction as carried out in this
paper.  We have proceeded in two steps because it is easier to see how
the reductions work this way, and because the intermediate stage,
which involves $T^*SU(2)$, is interesting in its own right, with
potential applications in quantum gravity.

Another possibility is to reduce the original problem by
$I_r=I'_r=J_r$, which means a set of twelve copies of the Hopf
fibration, resulting in a reduced phase space which is $(S^2 \times
S^2)^6$, in which the twelve angular momentum vectors $\Jvec_r$ and
$\Jvec'_r$, with fixed magnitudes, are coordinates.  In this case the
Ponzano-Regge phase can be expressed as a contour along 12-dimensional
Lagrangian manifolds in this 24-dimensional space, and the dihedral
angles appear as solid angles of wedges on spheres.  This phase space
can then be reduced again by $\Jvec_{\rm tot}$, which produces a phase
space that is closely related to the shape spaces, $\Shapespace$ and
$\Shapespacedot$. 

Thus far we have restricted attention to the Schl\"afli identity in
Euclidean $\Reals^3$. Schl\"afli's original proof treated the case
of the constantly curved three-sphere $S^3$, Schl\"afli (1858). In
fact, similar identities hold in each of the three-dimensional
constant curvature spaces, including hyperbolic three-space
$\mathbb{H}^3$, as well as in arbitrary dimensions; the metrical
proofs of Milnor (1994) and Alekseevskij, Vinberg and Solodovnikov
(1993) already mentioned treat the general case where the sectional
curvature $\kappa$ is variable.  If we parameterize this curvature
with $\kappa=-1,0,1$ for $\mathbb{H}^3$, $\mathbb{R}^3$ and $S^3$
respectively, then all three identities can be summarized in a single
formula 
	\begin{equation}
	2 \kappa d V = \sum_r J_r d \psi_r.
	\label{NEschlafli}
	\end{equation}
We believe that the symplectic method of proof described in this paper
can be generalized to an arbitrary constant curvature three
space. This is briefly outlined in the following paragraphs and will
be described more fully in a future publication.

Once again, we define a tetrahedron in each of the constant curvature
spaces $\mathbb{H}^3$ and $S^3$ through an ordered set of four
vertices.  We connect the vertices by geodesic arcs and view these arcs as
subsets of completely geodesic surfaces that make up the faces of the
tetrahedron. The curved space versions of the Schl\"afli identity
apply to these curved tetrahedra.

Just as a Euclidean tetrahedron emerges from the asymptotics of the
ordinary Wigner $6j$-symbol, a tetrahedron in spaces of constant
(positive or negative) curvature emerges from the asymptotics of the
$q$-deformed $6j$-symbol, an invariant in the representation theory of
the quantum group or Hopf algebra $\mathfrak{su}(2)_q$ (Chari and
Pressley 1994).  This much is clear already from the work of Mizoguchi
and Tada (1992) and of Taylor and Woodward (2005), both of whom used
one-dimensional WKB techniques along the lines of Schulten and Gordon
(1975ab). 

These asymptotics are important in quantum gravity. A number of works
have drawn connections between the $q$-deformed symbols and
discretizations of general relativity that include a cosmological
constant. In connection with loop quantum gravity Han (2011) and
Fairbairn and Meusberger (2011) have developed $q$-deformed spin foam
models and Dupuis and Girelli (2013), Bonzom \etal (2014a), Bonzom
\etal (2014b) and Dupuis \etal (2014) have developed further
connections between the geometry of $q$-deformation and the
cosmological constant.

A fully symplectic derivation of these asymptotics has yet to appear
and this is part of what we would like to achieve in our future work.
We will now outline some progress in this direction, speaking of the
case of real $q>0$, which corresponds to tetrahedral geometry in the
hyperbolic space $\mathbb{H}^3$.

The Hopf algebra $\mathfrak{su}(2)_q$ is generated by three operators,
$J_z$ and $J_\pm$, satisfying the commutation relations,
	\begin{equation}
	\eqalign{
	[J_z,J_\pm] &= \pm\hbar\, J_\pm, \cr
	[J_+,J_-] &= \hbar\, \frac{q^{J_z/\hbar} - q^{-J_z/\hbar}}
	{q^{1/2} - q^{-1/2}}}
	\label{su2qdef}
	\end{equation}
where in comparison to the usual formulas we have inserted factors of
$\hbar$, assuming that $J_z$, $J_\pm$ have dimensions of angular
momentum.  This algebra can be realized in terms of the raising and
lowering operators of a $q$-deformed harmonic oscillator, in a
generalization of the Schwinger-Bargmann representation of $SU(2)$, as
shown by Biedenharn (1989) and Macfarlane (1989).  We follow
Biedenharn's convention for $q$ in (\ref{su2qdef}).  The $q$-deformed
$6j$-symbol is constructed out of irreducible representations of this
algebra, labeled by six values of the quantum number $j$, in a manner
similar to that used to construct the ordinary $6j$-symbol out of six
irreps of $SU(2)$.  In fact, the $q$-deformed symbol approaches the
ordinary one as $q\to1$, just as the Hopf algebra (\ref{su2qdef})
approaches the ordinary Lie algebra $\mathfrak{su}(2)$ in the same limit.

To study the asymptotics of the representations of the algebra
(\ref{su2qdef}) we wish to take the limit $\hbar\to0$.  This limit
does not exist if $q$ is held fixed, because of the terms $q^{\pm
J_z/\hbar}$.  To obtain a well defined limit we follow Mizoguchi and
Tada (1992) and Taylor and Woodward (2005) by scaling $q$ along with
$\hbar$.  We do this by requiring that the ratio
	\begin{equation}
	J_0 = \frac{\hbar}{\log q}
	\label{J0def}
	\end{equation}
be held fixed in the limit $\hbar\to0$.  With this understanding the
limit $\hbar\to0$ converts the operator algebra (\ref{su2qdef}) into
the Poisson algebra
	\begin{equation}
	\eqalign{
	\{J_z,J_x\} &=J_y, \qquad \{J_z,J_y\} = -J_x, \cr
	\{J_x,J_y\} &= J_0 \sinh(J_z/J_0),}
	\label{JqPBs}
	\end{equation}
where $J_\pm=J_x\pm iJ_y$. In these Poisson bracket relations the operators
$(J_x,J_y,J_z)$ have been ``dequantized,'' that is, they are no longer
linear operators but rather functions on the phase space $\Phi$ of the
2-dimensional $q$-deformed harmonic oscillator.  We do not give the
definitions of these functions here but they are the dequantization of the
formulas given by Biedenharn (1989) and Macfarlane (1989) for the
$J$-operators in terms of the $q$-deformed raising and lowering
operators.  As for $J_0$, it plays the role of the curvature of the
hyperbolic space $\mathbb{H}^3$.  If we like we can set $J_0=1$,
corresponding to the standard case $\kappa=-1$ in (\ref{NEschlafli}),
or we can keep $J_0$ which is useful for studying the limit
$J_0\to\infty$ which is the nondeformed case $q=1$.

The definitions of $(J_x,J_y,J_z)$ provide a map from the phase space
$\Phi$ to $\Reals^3$, in which the $J$'s are coordinates (this is a
generalization of the Hopf map discussed below (\ref{IJdef})). The
Poisson bracket relations (\ref{JqPBs}) mean that this version of
$\Reals^3$ is a Poisson manifold.  In the nondeformed case ($q=1$),
the resulting space is $\mathfrak{su}(2)^*$, the dual of the Lie
algebra of $SU(2)$, otherwise ``angular momentum space.''  This is a
vector space, hence an Abelian group under addition.  In the deformed
case, the space with $(J_x,J_y,J_z)$ as coodinates is a non-Abelian
group, isomorphic to the subgroup of $SL(2,\Complexes)$ consisting of
upper triangular matrices with real diagonal elements.  We call this
group $B$; the identification of angular momentum space with matrices
in $B$ is specified by the matrix
	\begin{equation}
	b=\left(\begin{array}{cc}
	e^{-J_z/2} & -J_- \cr
	0 & e^{J_z/2}
	\end{array}\right)\in B,
	\label{bmatrix}
	\end{equation}
where we have set $J_0=1$.  Since these are elements of
$SL(2,\Complexes)$ they correspond to Lorentz transformations and thus
have an action on the unit mass shell in Minkowski $\Reals^4$, which
otherwise is a realization of the hyperbolic space $\mathbb{H}^3$. By
this action ``angular momentum space'' in the deformed case,
previously identified with $\Reals^3$, is seen to be diffeomorphic to
$\mathbb{H}^3$, which is a more useful identification.  The group
$B\subset SL(2,\Complexes)$ plays a role in the Iwasawa decomposition
of $SL(2,\Complexes)$ (Helgason 1978), in which an arbitrary element
of $SL(2,\Complexes)$ is factored uniquely into a product $bu$, where
$b\in B$ and $u\in SU(2)$.  Elements $b\in B$ can be used to identify
cosets in the space $SL(2,\Complexes)/SU(2)$, which otherwise is the
hyperbolic space $\mathbb{H}^3$.  The group $B \cong \mathbb{H}^3$ is
by (\ref{JqPBs}) and (\ref{bmatrix}) not only a group but a
Poisson manifold.  In this manner there appear the elements of a
Poisson-Lie group (Chari and Pressley 1994, Kosman-Schwarzbach 2004).

The comultiplication rule of the Hopf algebra (\ref{su2qdef}) can be
written in physicist's language as
	\begin{equation}
	\eqalign{
	J_z &= J_{1z} + J_{2z}, \cr
	J_\pm &= q^{-J_{1z}/2} \, J_{2\pm} +
	J_{1\pm} \, e^{J_{2z}/2},}
	\label{comultrule}
	\end{equation}
where we have set $\hbar=1$.  This is a generalization of the addition
of angular momenta, $\Jvec = \Jvec_1 + \Jvec_2$, in the nondeformed
case.  The significance of (\ref{comultrule}) is that if
$(J_{1z},J_{1\pm})$ and $(J_{2z},J_{2\pm})$ satisfy the Hopf algebra
(\ref{su2qdef}), then so do $(J_z,J_\pm)$, so that products of
representations of the Hopf algebra are also representations.  When
(\ref{comultrule}) is dequantized, we obtain
	\begin{equation}
	\eqalign{
	J_z &= J_{1z} + J_{2z}, \cr
	J_\pm &= e^{-J_{1z}/2}\, J_{2\pm} + J_{1\pm}\,e^{J_{2z}/2}}
	\label{Poissoncomult}
	\end{equation}
where we have set $J_0=1$.  This version of the comultiplication rule
has the property that if $\Jvec_1$ and $\Jvec_2$ satisfy the Poisson
algebra (\ref{JqPBs}), then so does $\Jvec$.  It also has the property
that if $\Jvec_1$ and $\Jvec_2$ correspond to group elements $b_1, b_2
\in B$, according to (\ref{bmatrix}), then $\Jvec$ corresponds to
group element $b=b_1b_2$.  In other words, the comultiplication rule
(\ref{Poissoncomult}) is the multiplication law for the group $B$ in
coordinates $\Jvec$ on $B$.

In particular, if $\Jvec=0$ in (\ref{Poissoncomult}), then $b_1b_2=1$,
or $b_2 = b_1^{-1}$.  By letting elements of $B$ act on the origin in
$\mathbb{H}^3$ those elements can be used to label points of
$\mathbb{H}^3$, and a relation such as $b_1b_2=1$ can be interpreted
geometrically as the tracing and retracing a geodesic line segment in
$\mathbb{H}^3$.  That is, it becomes a $q$-deformed version of a
diangle condition.  (Our notation here is slightly different from that
used in Sec.~\ref{ABmanifolds} for the diangle condition:
$\Jvec_1$ and $\Jvec_2$ here correspond to $\Jvec$ and $\Jvec'$
there.)  The length of the line segment can be interpreted as the
value of the $q$-deformed version of the Casimir function, the
generalization of the function $I$ in (\ref{IJdef}), for either
$\Jvec_1$ or $\Jvec_2$.  Combining the diangle condition $\Jvec=0$
with the Casimir conditions gives the specification of a Lagrangian
manifold in the $q$-deformed version of the phase space
$\Phi_{2j}=\Phi\times\Phi$, and taking the six-fold product of this
gives us the $q$-deformed version of the $B$-manifold, a Lagrangian
submanifold of $\Phi_{12j}$.  

Comultiplication is associative but not commutative.  Taking the
coproduct of three copies of the Poisson algebra (\ref{JqPBs}) gives
	\begin{equation}
	\eqalign{
	J_z &= J_{1z} + J_{2z} + J_{3z}, \cr
	J_\pm &= e^{-(J_{1z}+J_{2z})/2}\, J_{3\pm} +
	         e^{-J_{1z}/2}\, J_{2\pm}\, e^{J_{3z}/2} +
	         J_{1\pm} \, e^{(J_{2z}+J_{3z})/2}.}
	\label{3Jcomult}
	\end{equation}
Now the condition $\Jvec=0$ corresponds to $b_1b_2b_3=1$ when
$\Jvec_1$, $\Jvec_2$ and $\Jvec_3$ are mapped into group elements by
(\ref{bmatrix}), and this in turn can be interpreted geometrically as
a triangle condition in $\mathbb{H}^3$.  When the values of the three
Casimirs (that is, the lengths of the edges of the triangle) are
fixed, we obtain a Lagrangian manifold in the product space
$\Phi\times\Phi\times\Phi$.  And when four triangle conditions are
combined in $\Phi_{12j}$ we obtain the $q$-deformed version of the
$A$-manifold, a Lagrangian submanifold of $\Phi_{12j}$. 

Thus we obtain the basic geometrical picture for the asymptotics of
the $q$-deformed $6j$-symbol.  The $A$- and $B$- manifolds have
intersections which correspond geometrically to a tetrahedron in
$\mathbb{H}^3$, and the phase of the asymptotic expression is the
integral of the symplectic form from one component of the
intersection, along the $B$-manifold to the other component, and then
back along the $A$-manifold to the first component, with a final
motion along the intersection to the initial point.  The details are
more complicated than in the nondeformed case mainly because the group
$B$ is non-Abelian, but they follow the outline presented in this
paper for the nondeformed case.  We will report on this calculation
and its relation to the Schl\"afli identity more fully in future
publications.

\ack

HMH acknowledges support from the National Science Foundation (NSF)
International Research Fellowship Program (IRFP) under Grant
No. OISE-1159218.

\section*{References}
\begin{harvard}

\item[] Abraham R and Marsden J E 1978 {\it Foundations of Mechanics}
(Reading, Massachusetts:  Benjamin/Cummings)

\item[] Alekseevskij D V, Vinberg V and Solodovnikov A S 1993 {\it
Encyclopedia Math. Sci.} {\bf 29} 1

\item[] Aquilanti V, Haggard H M, Littlejohn R G and Yu L 2007
{\it J Phys A} {\bf 40} 5637

\item[] Aquilanti V, Haggard H M, Hedeman A, Jeevanjee N, Littlejohn 
R G and Yu L 2012 {\it J. Phys. A} {\bf 45} 065209

\item[] Arnold V I 1989 {\it Mathematical Methods of Classical
Mechanics} (New York: Springer-Verlag)

\item[] Bahr B and Dittrich B 2009 {\it Class Quantum Grav} {\bf 26} 225011

\item[] \dash 2010 {\it New Journal of Physics} {\bf 12} 033010

\item[] Balazs N L and Jennings B K 1984 {\it Phys. Reports} {\bf 104}
347

\item[] Bargmann V 1962 {\it Rev. Mod. Phys.} {\bf 34} 829

\item[] Barrett J W and Steele C M 2003 {\it Class Quantum Grav} {\bf
20} 1341

\item[] Bates S and Weinstein A 1997 {\it Lectures on the Geometry of
    Quantization} (Providence, Rhode Island: American Mathematical
  Society)

\item[] Ben Geloun J and Gurau R 2011 {\it Ann Inst Henri Poincar\'e}
{\bf 12} 77

\item[] Berry M V 1977 {\it Phil. Trans. Roy. Soc.} {\bf 287} 237

\item[] \dash 1984 {\it Proc Roy Soc Lond A} {\bf 392}, 45

\item[] Biedenharn L C and Louck J D 1981 {\it The Racah-Wigner 
Algebra in Quantum Theory} (Reading, Massachusetts: Addison-Wesley)

\item[] Biedenharn L C 1989 {J Phys A} {\bf 22} L873

\item[] Blau M and Thompson G 1995 preprint hep-th 9501075

\item[] Bonzom V, Dupuis M, Girelli F and Livine E R (2014a) preprint 
gr-qc 1402.2323

\item[] Bonzom V, Dupuis M, Girelli F (2014b) preprint gr-qc 1403.7121

\item[] Carfora M and Marzuoli A 2012 {\it Quantum Triangulations}
(New York: Sprnger)

\item[] Cushman R H and Bates L 1997 {\it Global Aspects of Classical 
Integrable Systems} (Basel: Birkh\"auser Verlag)

\item[] Dittrich B, Freidel L and Speziale S 2007 {\it Phys Rev D}
{\bf 76} 104020

\item[] Duistermaat J J and Heckman G J 1982 {\it Invent Math} {\bf
69} 259

\item[] Dupuis M and Girelli F 2013 {\it Phys Rev D} {\bf 87} 121502(R)

\item[] Dupuis M, Girelli F and Livine E (2014) preprint gr-qc 1403.7482

\item[] Echeverr\'\i a-Enr\'\i quez A, Mu\~noz-Lecanda M C,
  Rom\'an-Roy N and Victoria-Monge C 1999 preprint math-ph 9904008

\item[] Edmonds A R 1960 {\it Angular Momentum in Quantum Mechanics}
(Princeton: Princeton University Press)

\item[] Freidel L and Speziale S 2010 {\it Phys Rev D} {\bf 82} 084040

\item[] Guillemin V, Jeffrey L and Sjamaar R 2002 {\it Transformation
Groups} {\bf 7} 155

\item[] Haggard H M and Littlejohn R G 2010 {\it Classical and Quantum
Gravity} {\bf 27} 135010

\item[] Han M 2011 {\it J Math Phys} {\bf 52} 072501

\item[] Helgason, S 1978 {\it Differential Geometry, Lie Groups and
Symmetric Spaces} (New York: Academic Press)

\item[] Holm D D 2011 {\it Geometric Mechanics, part II: Rotating,
    Translating and Rolling} (London: Imperial College Press)

\item[] Chari V and Pressley A 1994 {\it A Guide to Quantum Groups} 
(Cambridge: Cambride University Press)

\item[] Kirillov A A 1976 {\it Elements of the Theory of
Representations} (New York: Springer-Verlag)

\item[] Kneser H 1936 {\it Deutsche Math.} {\bf 1} 337

\item[] Kosman-Schwarzbach Y 2004 {\it Lect Notes Phys} {\bf 638} 107

\item[] Littlejohn R G 1990 {\it  J. Math.\ Phys.} {\bf 31} 2952

\item[] Littlejohn R G and Reinsch M 1995 {\it Phys Rev A} {\bf 52}
2035

\item[] \dash 1997 {\it Rev Mod Phys} {\bf 69} 213

\item[] Littlejohn R G and Yu L 2009 {\it J. Phys. Chem. A} {\bf 113} 
14904

\item[] Livine E R and Tambornino J 2011 preprint gr-qc 1105.3385

\item[] Livine E R and Oriti D 2003 preprint gr-qc 0302018

\item[] Luo F 2008 {\it Commun. Contemp. Math.} {\bf 10} 835

\item[] Macfarlane 1989 {\it J Phys A} {\bf 22} 4581

\item[] Marsden J E and Ratiu T 1999 {\it Introduction to Mechanics and
  Symmetry} (New York: Springer-Verlag)

\item[] Messiah A 1966 {\it Quantum Mechanics} (New York: John Wiley)

\item[] Milnor J 1994 {\it Collected Papers} v.~1 (Houston, Texas: 
Publish or Perish)

\item[] Misner C W, Thorne K S and Wheeler J A 1973 {\it Gravitation}
  (New York: W. H. Freeman and Company)

\item[] Mizoguchi~S and Tada~T 1992 {\em Phys. Rev. Lett.} {\bf 68} 1795

\item[] Narasimhan R S and Ramadas T R 1979 {\it Commun Math Phys} 
{\bf 67} 121

\item[] Ozorio de Almeida Alfredo M 1998 {\it Phys. Reports} {\bf
295} 265

\item[] Ponzano G and Regge T 1968 in {\it Spectroscopy and Group
Theoretical Methods in Physics} ed F Bloch \etal\ (Amsterdam:
North-Holland) p~1

\item[] Regge T 1961 {\it Il Nuovo Cimento} {\bf 19}, 558

\item[] Regge T and Williams R M 2000 {\it J. Math. Phys.} {\bf 41} 3964

\item[] Rivin I and Schlenker J-M 2008 preprint math 0001176

\item[] Roberts J 1999 {\it Geometry and Topology} {\bf 3} 21

\item[] Schulten K and Gordon R G 1975a {\it J. Math. Phys.} {\bf 16} 1961

\item[] \dash 1975b {\it J. Math. Phys.} {\bf 16} 1971

\item[] Schl\"afli L 1858 {\it Quart. J. Pure Appl. Math.} {\bf 2} 269

\item[] Schwinger J 1952 {\it On Angular Momentum} U.S. Atomic Energy
Commission, NYO-3071, reprinted in Biedenharn L C and van Dam H 1965
{\it Quantum Theory of Angular Momentum} (New York: Academic Press)

\item[] Sforza G 1907 {\it Atti della Societ\`a di Naturalisti e
Matematici di Modena} {\bf 9} 1

\item[] Simms D J and Woodhouse N M J 1977 {\it Lectures on Geometric
Quantization} (New York: Springer Verlag)

\item[] Souam R 2004 {\it Diff Geom Appl} {\bf 20} 31

\item[] Stone M 1989 {\it Nucl Phys} {\bf B314} 557

\item[] Taylor Y U and Woodward C T 2005 {\it Selecta Math (N.S.)}
{\bf 11} 539

\item [] Thiemann T 2007 {\it Modern Canonical Quantum General 
Relativity} (Cambridge: Cambridge University Press)

\item[] Wigner E P 1959 {\it Group Theory} (Academic Press, New York)

\item[] Yakut A T, Savas M and Kader S 2009 {\it Geom Dedicata} {\bf
138} 99

\end{harvard}
\end{document}